\documentclass[12pt]{article}

\usepackage{chemformula}
\usepackage[T1]{fontenc}
\usepackage{subcaption}
\usepackage[percent]{overpic}
\usepackage{amsmath}
\usepackage{amssymb}
\usepackage{hyperref}
\usepackage{xcolor}
\usepackage[a4paper, left=1in, right=1in, top=1in, bottom=1in]{geometry}
\usepackage{float}

\title{Bulk photovoltaic effect in ferroelectric and antiferroelectric phases of antimony sulphoiodide investigated by means of ab-initio simulations}

\author{
    Giuseppe Cuono\textsuperscript{1}, 
    Subhadeep Bandyopadhyay\textsuperscript{1},
    Andrea Droghetti\textsuperscript{2}, 
    Silvia Picozzi\textsuperscript{3,1}
}

\date{}

\begin{document}

\maketitle

\begin{center}
\textsuperscript{1}Consiglio Nazionale delle Ricerche (CNR-SPIN), Unit\'a di Ricerca presso Terzi c/o Universit\'a “G. D’Annunzio”, 66100 Chieti, Italy. \\

\textsuperscript{2}Department of Molecular Sciences and Nanosystems, Ca’ Foscari University of Venice, via Torino 155, 30170, Venice-Mestre, Italy. \\

\textsuperscript{3}Department of Materials Science, University Milan-Bicocca, 20125 Milan, Italy.\\
E-mail: giuseppe.cuono@spin.cnr.it, subha.7491@gmail.com, andrea.droghetti@unive.it, silvia.picozzi@unimib.it
\end{center}

\begin{abstract}
We employ first-principles calculations to investigate the ferroelectric properties and the bulk photovoltaic effect (BPVE) of antimony sulfur iodide (SbSI). The BPVE enables direct sunlight-to-electricity conversion in homogeneous materials and, in ferroelectric compounds, can be tuned via an electric field controlling the polarization. However, most ferroelectrics are oxides with large band gaps exceeding the energy of visible light, thereby limiting their photovoltaic performance. SbSI, featuring a visible-range band gap, combines remarkable photovoltaic capabilities with a spin-textured band structure, coupling charge and spin degrees of freedom. Our calculations predict ferroelectric and antiferroelectric phases with comparable band gaps but distinct spin textures, relevant for spintronics applications. The BPVE is driven by the linear and circular photogalvanic effects, exhibiting high photoconductivities under visible light. Furthermore, it serves as a diagnostic tool to identify the material phase, with the circular photogalvanic effect reflecting spin texture changes. Thanks to its multifunctional properties, SbSI emerges as a promising candidate for solar energy conversion and advanced electronics, with potential applications extending to spintronics.
\end{abstract}

\maketitle

\section{Introduction}


The bulk photovoltaic effect (BPVE) enables the conversion of solar energy into electrical power within homogeneous systems \cite{Dai23,Fridkin01,VonBaltz81,Sipe00}. Unlike the conventional photovoltaic effect, which relies on a built-in potential at material interfaces (e.g., p–n junctions) to separate photogenerated charge carriers, the BPVE allows for simpler device architectures where the entire bulk material contributes to photocurrent generation. Furthermore, the BPVE has been reported to produce above-band-gap open-circuit voltages, as charge carriers can be collected before thermalization, potentially exceeding the Shockley–Queisser limit that constrains conventional solar cells \cite{Shockley60,Spanier16}. However this claim remains under debate, as other factors may limit energy conversion efficiency \cite{Kirk17,Pusch23}.

The primary contributions to the BPVE are the shift and injection currents \cite{Ma23,Sipe00,Ahn20,Wang20}, both second-order responses to the incident light field.
The shift current \cite{Kral00} is an interband phenomenon, which can be interpreted as a real-space displacement of electrons upon light excitation \cite{Sipe00,Azpiroz18,Puente23}. In contrast, the injection current arises from asymmetries in the velocity distribution of excited carriers \cite{Dai23} due to non-reciprocal band structures \cite{Zhang2019,stavric2025} or processes like asymmetric carriers photoexcitation, recombination, and momentum relaxation \cite{Dai21}. Recently, either the shift current, the injection current, or both have been investigated experimentally \cite{Auston72,Asnin,  Cote02, McIver2012,Nakamura2017, Rees2020,Yuan2014, Niu23,Moldavskaya23, Urakami24} as well as theoretically \cite{Ivchenko78, Nastos10, deJuan2017,Tiwari22,Pal21,Dey24,Zhang18,Sadhukhan25,Zhang19b,Zhang19,Sandhukan20,Sandhukan21,Sadhukhan21b,Le20,Stavric25} across numerous systems.

Within a single-particle framework (i.e., considering only band structure effects), symmetry considerations allow these currents to be further classified based on whether they are driven by linearly or circularly polarized light. Specifically, in non-magnetic materials where time-reversal symmetry is preserved, linearly polarized light induces the shift current (linear photogalvanic effect), whereas circularly polarized light generates the injection current (circular photogalvanic effect), whose direction depends on the light’s helicity \cite{Wang20,Puente23}.
The linear photogalvanic effect occurs in noncentrosymmetric materials, and more precisely, in materials with piezoelectric point groups, while the circular photogalvanic effect is seen in a further subset of these systems, known as gyrotropic materials (out of the 32 point groups, 21 are noncentrosymmetric; among these, 20 point groups exhibit piezoelectric properties, and 18 of those are also gyrotropic). 

Among the materials exhibiting both photogalvanic effects, ferroelectric (FE) compounds with spontaneous electrical polarization have attracted significant interest \cite{Dai23,Koch76, Young12,Young12b,Butler15,Tan16,Pal21},
as the photocurrent direction can be controlled by reversing the polarization via an electric field. Despite that, the most studied FE materials, such as the perovskite oxides LiNbO$_3$, BaTiO$_3$ \cite{Koch76, Young12}, BiFeO$_3$ \cite{Young12b} and PbTiO$_3$ \cite{Young12}  have band gaps greater than 3 eV, limiting their ability to capture sunlight to the ultraviolet (UV) range, which accounts for only about 3.5\% of the total solar spectrum. Ideal materials for studying the BPVE should possess narrower band gaps and strong absorption in the visible spectrum \cite{Cook17}.

As an alternative to FE oxide materials, group-IV monochalcogenide monolayers, such as GeS, GeSe, SnS, and SnSe, have been proposed due to their smaller band gaps and high carrier mobilities \cite{Rangel17}. 
While these monochalcogenides are centrosymmetric in their bulk form, they lack inversion symmetry in their monolayer configurations, which allows for the emergence of spontaneous electrical polarization and large shift  \cite{Rangel17} and injection currents \cite{Panday19} under visible light. Moreover, first-principles calculations predict that these currents are enhanced owing to the reduced dimensionality of the materials \cite{Rangel17}.
Despite that, experimental measurements on non-centrosymmetric layer-stacked SnS have reported much lower photocurrents than expected \cite{Chang23,Nanae2024}. 
This discrepancy has been attributed to the presence of multiple ferroelectric domains with opposing spontaneous polarizations in the samples, which reduces the overall shift current by averaging out the individual domain contributions \cite{Chang23}.

Among the FE materials, antimony sulfoiodide (SbSI) is another promising alternative to oxides owing to its visible-range band gap \cite{Madelung04}, equal to approximately 1.9 eV. 
SbSI belongs to the class of multifunctional materials known as FE Rashba-Dresselhaus semiconductors \cite{DiSante13,Picozzi14}. 
In these materials, strong spin-orbit coupling (SOC) is combined with ferroelectricity resulting in spin textured electronic bands characterized by a large Rashba-Dresselhaus spin splitting that can be controlled by applying an electric field \cite{Picozzi14}. This effect can find applications in novel device concepts for spintronic applications \cite{Varotto2021}. Thus, FE Rashba-Dresselhaus semiconductors possess the essential characteristics for the BPVE, which, moreover, can potentially be further integrated with electronic and spin transport functionalities.

The discovery of ferroelectricity in SbSI dates back to 1962, with a reported spontaneous polarization of about 25 $\mu$C/cm$^2$ at 0$^\mathrm{o}$ C \cite{Fatuzzo62}.  It was later found that SbSI also exhibits an antiferroelectric (AFE) phase in the temperature range 295 K $<\mathrm{T}<$ 410 K, above which the material transitions to the paraelectric phase \cite{Audzijonis98,Audzijonis08}.
In the FE phase, SbSI exhibits high photoconductivity \cite{Nitsche60} and nonlinear optical behavior \cite{Harbeke63}. The shift current has been experimentally measured in several studies \cite{Ogawa17,Sotome19,Nakamura18,Mistewicz19}, yielding an extrapolated photoconductivity of $\sim10$ $\mu$A/V$^2$ - one of the largest reported values for a FE semiconductor. 
While significant progress has been made in studying this phenomenon, a comprehensive theoretical analysis that derives all components of the photoconductivity tensor, evaluates their relative contributions across the visible spectrum, and explores the upper bounds of photocurrent generation remains to be fully developed. In addition, the AFE phase has been comparatively less explored, despite its stability near room temperature.

In this paper, we use first-principles calculations based on Density Functional Theory (DFT) to fill this knowledge gap. Specifically, we begin by comparing the FE and AFE structures, elucidating the mechanisms that stabilize these phases. We then demonstrate that the phase transition is accompanied by a significant modification of the spin texture in the band structure - a feature that may be of interest for spintronic applications. Finally, we analyze the BPVE, accurately computing both the shift and circular injection photoconductivities by deriving all symmetry-allowed tensor components and estimating their magnitudes across the visible energy range.

Three main findings emerge from our study. First, the photocurrent can reach very large peak values, significantly exceeding those observed in experiments to date. This points to untapped potential for enhancing the material’s photovoltaic performance. 
Second, we identify how the shift and circular photocurrents evolve across the FE-to-AFE transition. These insights can be potentially important for setting up measurements capable of distinguishing between the two phases.
Third, we show how circular injection current measurements can track the change in spin texture from the FE phase to the AFE phase, revealing a rich interplay among structural, optical, and spintronic properties.

The paper is organized as follows. The computational details are provided in Section~\ref{sec.comp_details}. The results are presented in Section~\ref{sec.results}, where we describe the structural properties in Subsection~\ref{sec.structure}, estimate the macroscopic polarization in Subsection~\ref{sec.pol}, analyze the band structure in Subsection~\ref{sec.bands}, and discuss the photoconductivities in Subsection~\ref{sec:BPVE}. Finally, we summarize our conclusions in Section~\ref{sec:conclusions}.

\begin{figure*}[t!]
\centering
\includegraphics[width=\textwidth]{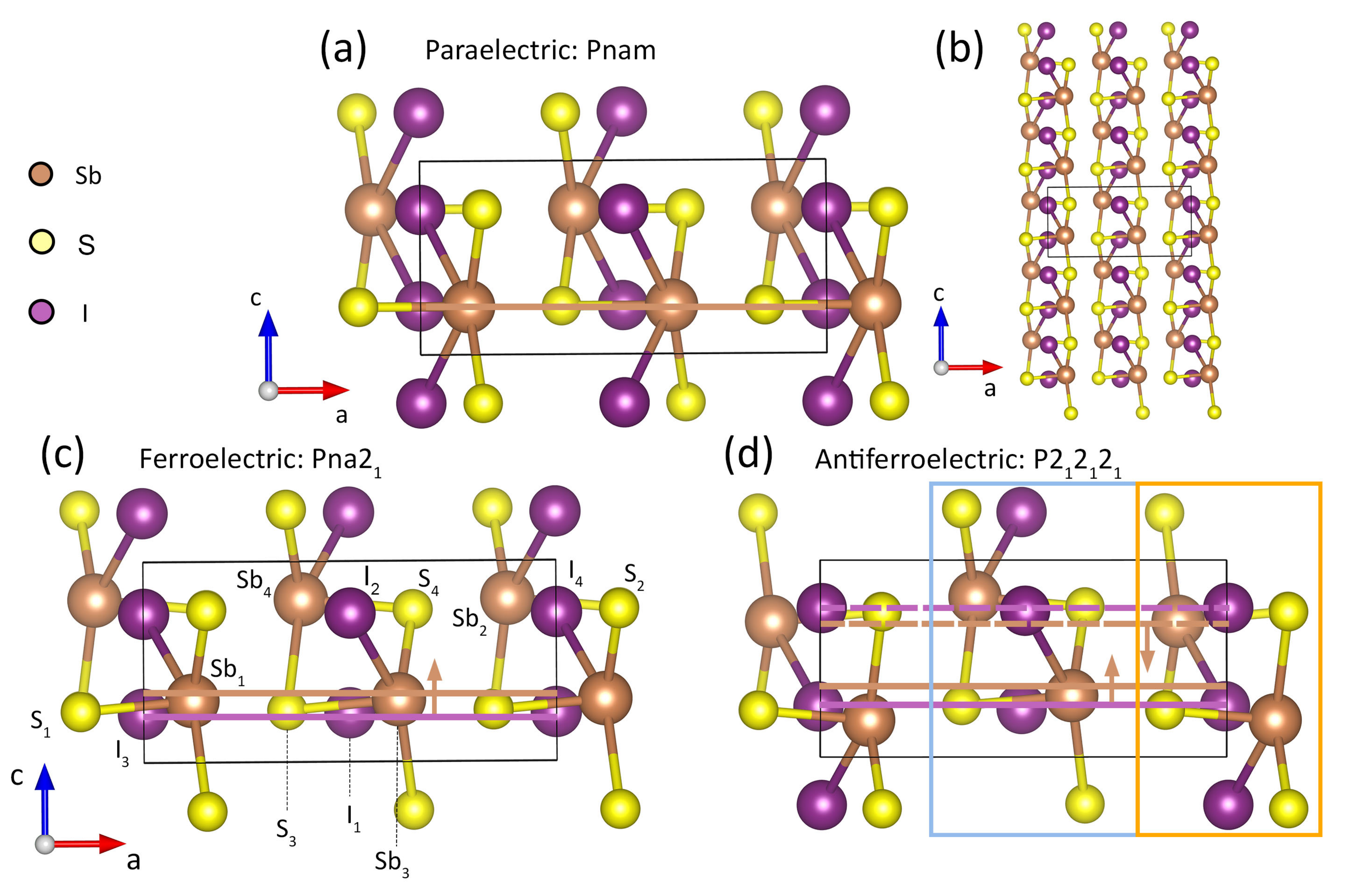}
\caption{Crystal structure of SbSI in the PE (space group Pnam), FE (space group Pna2$_1$) and AFE (space group P2$_1$2$_1$2$_1$) phases. (a) In the PE configuration, the Sb, S and I ions occupy the same position along the $c$ axis (see brown line). The polarization is zero. (b) The system is composed of zig-zag chains propagating along the $c$ axis. (c) In the FE phase, the Sb sub-lattice is shifted with respect to the anions sub-lattice along the $c$ axis and FE polarization appears. In this configuration the cations are shifted in the same direction in both chains, as indicated by the brown arrow. The cation and anion sub-lattices have different z coordinates (see brown and purple lines). (d) In the AFE phase, the cations of one chain shift in one direction, while the cations of the other chain shift in the opposite direction, as indicated by brown arrows. The blue and orange rectangles highlight the two different chains. The solid brown and purple lines indicate the positions along $c$ of the cations and anions in one chain, while the dashed brown and purple lines in the other chain. In the AFE configuration the polarization is zero; however,  in both the FE and AFE phases the inversion symmetry is broken.  }\label{structure}
\end{figure*}

\section{Computational details}\label{sec.comp_details}

DFT calculations are performed by using the projector-augmented
wave method, as implemented in the VASP package \cite{Kresse93,Kresse96,Kresse96b}. The Perdew-Burke-Erzenhof (PBE) \cite{Perdew96} generalized gradient approximation (GGA) is used for the exchange-correlation functional in most of our simulations, while the Heyd-Scuseria-Ernzerhof 2006 (HSE06) hybrid density functional approach \cite{Krukau06,Paier06} is employed to correct for the short-coming of the GGA in describing the energetics of the system and the size of the band gap. The van der Waals (vdW) dispersion energy is included in the structural relaxation using the DFT-D3 method of Grimme {\it et al.} \cite{Grimme10}. The plane-wave energy cutoff is set to 400 eV.
The atomic positions are optimized until the residual forces are smaller than 0.01 eV {\AA}$^{-1}$. The calculation of the macroscopic electric polarization is performed within the modern theory of polarization through the Berry phase method \cite{Smith93}. For the geometry optimizations, the Brillouin zone is sampled with 4 $\times$ 4 $\times$ 8 Monkhorst-
Pack $k$-grid centered at the $\Gamma$ point. This grid is refined to 8 $\times$ 8 $\times$ 12 in the self-consistent calculations at the optimized geometries, since the photoresponses are highly sensitive to the used $k$-mesh. 

The linear and circular photogalvanic effects are described in terms of their photoconductivity tensors, using the approach developed by Azpiroz {\it et al.} \cite{Azpiroz18} and Puente-Uriona {\it et al.} \cite{Puente23}. This approach is based on interpolation of the DFT band structure using maximally-localized Wannier functions \cite{Marzari97,Mostofi08}, as implemented in the {\sc Wannier90} package \cite{Pizzi2020}. 
The $s$ and $p$-orbitals of Sb, S, and I ions are considered, resulting in a total of 96 bands when spin-orbit coupling (SOC) is included. For  bulk systems, a 100 $\times$ 100 $\times$ 100 $k$-grid is used together with the adaptive broadening scheme described in Ref. \cite{Yates07} to cope with the Dirac delta functions. For the single-chain system, a 1 $\times$ 1 $\times$ 100 $k$-grid is employed, with a fixed broadening of 0.02 eV to better handle the van Hove singularities, following the method of Azpiroz {\it et al.} \cite{Azpiroz18}. With these $k$-grids, the photoconductivities are well converged, and no significant differences in the results are observed when the grid density is further increased.
SOC is included in all calculations of the band structure and photocurrents, while it is neglected in the geometry optimizations, as it is not expected to significantly affect the structural parameters. 

Symmetry mode analysis was performed using the ISODISTORT software \cite{Isodistort, Isodistort1} to identify the atomic distortions associated with the AFE and FE phases, using the paraelectric (PE) Pnam phase as the reference structure.

\section{Results and discussion}\label{sec.results}
\subsection{Structural properties}\label{sec.structure}

We begin by presenting the structural and electronic properties of SbSI. While the FE phase has been previously investigated \cite{Amoroso16}, the AFE phase has received comparatively little attention, and a comprehensive theoretical comparison between the two phases is still lacking. The results will also lay the groundwork for understanding the photoresponse.

In the paraelectric (PE) phase, SbSI has an orthorhombic crystal structure, which belongs to the centrosymmetric space group Pnam \cite{Kikuchi67}, with four formula units (i.e., 12 atoms) in the primitive unit cell [Fig. \ref{structure}(a)].
The structure consists of zigzag chains propagating along the $c$-axis (assumed parallel to the $z$ Cartesian axis) which are weakly bonded by van der Waals interactions [Fig. \ref{structure}(b)]. The primitive unit cell contains two zigzag chains. Within each chain, the atoms are grouped in parallel layers in the $xy$ plane, and the bonds are characterized by a high degree of covalency. However, for a simplified understanding, one can  describe a chain in terms of ionic bonds, assigning oxidation states of Sb$^{3+}$, S$^{2-}$ and I$^{1-}$ to the atoms.

In the PE phase, both cations and anions are placed at the same $z$ coordinate. In contrast, in the FE phase, a cooperative displacement of Sb atoms results in a shift of the cation sub-lattice along the $c$-axis relative to the anion I and S  sub-lattices, as shown in Fig. \ref{structure} (c). 
By performing a symmetry mode analysis, we identify this atomic distortion as the irreducible $\Gamma_4^-$ mode relative to the PE Pnam structure. The $\Gamma_4^-$ mode lowers the symmetry of the system to Pna2$_1$, while preserving the orthorhombic character.
The Sb ions play a primary role by displacing more than the other atoms with respect to their centrosymmetric positions, breaking inversion symmetry and creating a polarization along the $c$-axis \cite{Amoroso16}. This behavior can be attributed to the presence of a Sb$^{3+}$ lone pair, extending in the empty region between the chains and tilting along the $c$ axis in the FE phase so to minimize the Coulomb repulsion with the other bonding electrons \cite{Amoroso16}.


The AFE phase crystallizes in the orthorhombic P2$_1$2$_1$2$_1$ structure, as reported in experiments \cite{Audzijonis98, Audzijonis08}. Our symmetry mode analysis identifies the $\Gamma_1^-$ distortion relative to the Pnam PE phase as the driving mechanism [see Section S1 in the Supplemental Material (SM) \cite{Supplemental}]. This $\Gamma_1^-$ mode corresponds to an antipolar distortion of the cations (along the $c$-axis) between adjacent atomic chains, as shown in Fig. \ref{structure}(d). As a result, each chain exhibits polarization, but the polarizations of neighboring chains are oriented in opposite (antiparallel) directions, leading to a net zero polarization for the overall system. We note that, in addition to the $\Gamma_1^-$ mode, the fully relaxed P2$_1$2$_1$2$_1$ structure also includes a $\Gamma_1^+$ atomic distortion and related $\eta_{\Gamma_1^+}$ strain, which, however, do not break the Pnam symmetry.

The FE distortion can not be reproduced by means of DFT within the standard GGA functional because the structural optimization tends to converge to the PE structure, as already reported in previous studies \cite{Amoroso16}. To overcome this problem, we perform ionic relaxations using the hybrid functional HSE06 \cite{Paier06}, by also taking into account the Van der Waals interaction among the chains. The HSE06 functional is known to effectively capture the delicate balance between electronic properties and lattice distortions \cite{Franchini09, Droghetti10}. Similarly to the FE phase, the HSE functional is also found to be crucial for relaxing the AFE phase into the P2$_1$2$_1$2$_1$ structure.

In our HSE calculations, the FE configuration is found to be lower in energy than the PE and AFE configurations by 4.20 meV and 2.59 meV per formula unit, respectively; in other words, the AFE configuration is the second lowest in energy. Therefore, our results reproduce the stability ordering of the three phases observed in experiments. Moreover, since the FE and AFE phases are close in energy, this suggests that the material is likely to exhibit antiferroelectricity at finite temperatures \cite{Subhadeep_BiNiO3}.

The optimized lattice parameters and atomic positions in the PE, FE and AFE phases are reported in detail in the Section S2 of the SM.
The obtained values are in good agreement with experiments. For example, in the FE phase, our optimized lattice parameters are $a=8.56$ {\AA}, $b=10.10$ {\AA} and $c=4.16$ {\AA}, while the experimental ones are respectively 8.49 {\AA}, 10.10 {\AA} and 4.16 {\AA} \cite{Donges50}. The lattice parameters only slightly differ in the AFE and FE phases.
Notably, in both these phases, the chains tend to elongate along the $c$-axis with respect to the PE geometry. In fact, we find that the $c$  lattice parameter increases from 4.10 {\AA} in the PE phase to 4.16 {\AA}, and 4.17 {\AA} in the FE and AFE phases, respectively. This points to a ``strain-polarization'' coupling: while polarization develops, the $c$-axis elongates, in close similarity to what happens for standard ferroelectrics, such as P4mm BaTiO$_3$. In the PE phase, the atom labeled Sb$_3$ in Fig. \ref{structure}(c) is positioned at the same $z$ coordinate as the I$_{1}$ and S$_{3}$ atoms. In contrast, in both the FE and AFE phases, Sb$_3$ is displaced along the $c$ axis relative to I$_{1}$ and S$_{3}$ by $\Delta z_{\mathrm{Sb}_{3}-\mathrm{I}_{1}}=0.29$~\AA~ and $\Delta z_{\mathrm{Sb}_{3}-\mathrm{S}_{3}}=0.21$~\AA. These values, which quantify the polar distortion, are slightly overestimated but in reasonable agreement with the experimental values for the FE phase, which are 0.21~\AA~and 0.17~\AA \cite{Oka66}. This demonstrates the good performance of HSE calculations in predicting the FE structure of the material.

\begin{figure}[t!]
\centering
\includegraphics[width=10cm,angle=0]{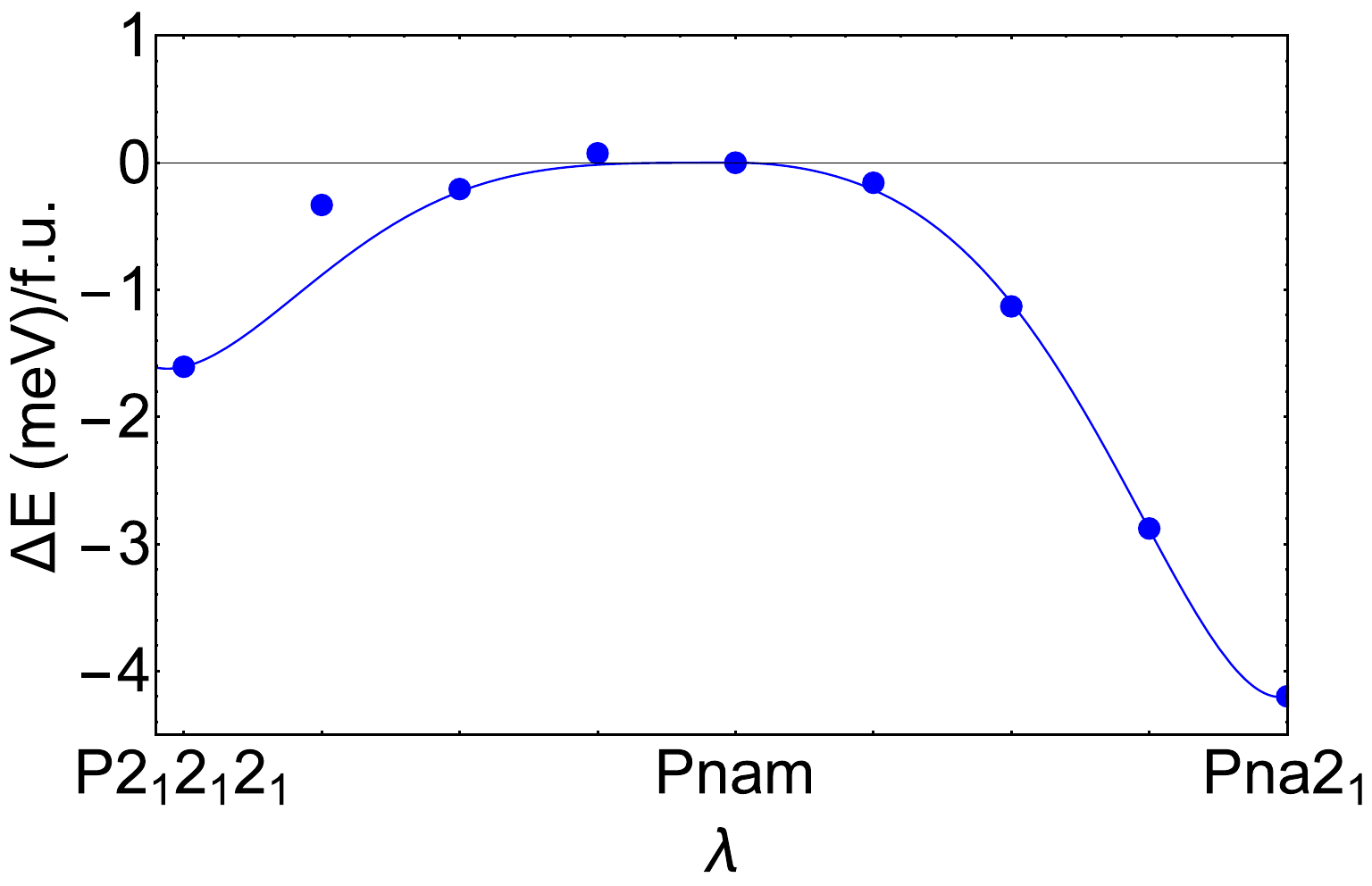}
\caption{Energy path from the AFE (P2$_1$2$_1$2$_1$) to the FE phase (Pna2$_1$), through the PE configuration (Pnam), built by means of linear interpolation of lattice parameters and coordinates. The curve follows Landau theory and represents an expansion of the energy difference in even powers of the generalized parameter $\lambda$, up to the sixth order.}\label{Energy_path}
\end{figure}

\subsection{Electrical polarization}\label{sec.pol}
The macroscopic electrical polarization, $\mathbf{P}$, is calculated within the modern theory of polarization through the Berry phase method \cite{Smith93}. We build an adiabatic path from the optimized PE to FE and AFE structures by linearly interpolating both lattice parameters and atomic positions, labeled by a generalized coordinate $\lambda$, as depicted in Fig. \ref{Energy_path}. The parameter runs from -1 (AFE structure with space group P$2_1$2$_1$2$_1$) through 0 (PE structure with Pnam space group), and up to 1 (FE structure with Pna2$_1$ space group). The resulting energy profile shows a characteristic (asymmetric) double-well shape, with an energy difference of 4.20 meV per f.u. between the FE and PE phase, and a lower energy difference of 2.59 meV between the AFE and the PE phase, as anticipated in the previous subsection.
We note that our estimated FE–PE energy difference is slightly larger than the value of approximately 1.5 meV/f.u. reported in Ref. \cite{Amoroso16}.  In that work, only changes in atomic positions were considered keeping the lattice parameters unchanged. In our simulations, we calculate energy difference between fully optimized PE and FE phases, where we considered both atomic and lattice relaxations.


The macroscopic polarization in the FE phase lies along the $c$-axis, as expected from symmetry, with a calculated magnitude of $P\simeq31$ (30) $\mu$C/cm$^2$ with the HSE06 (GGA) functional, in good agreement with  experimental results \cite{Fatuzzo62} and previous DFT calculations \cite{Amoroso16}. Notably, for fixed structures, the polarization value is very similar when using either the GGA or HSE functional. Therefore, even though
the GGA fails in estimating the correct energetics of the system and in the geometry optimization of the FE state, it performs well for the ferroelectric properties.

The macroscopic polarization in the AFE phase is zero. However, we can estimate the polarization of a single chain approximately by computing the tensor of the Born effective (or dynamical) charges (BEC) \cite{Ghosez98,Gonze97}, $Z^{*}_{ab}$ with $a,b=x,y,z$. The contribution  of each atom to the polarization can be defined as the product of the BEC and the displacements of the atom in all three Cartesian directions relative to their positions in the PE configuration \cite{Spaldin2012}. The polarization of a single chain is then obtained by summing over the contributions of all the atoms in the chain. In Table \ref{BEC_SbSI}, we provide the components of the BEC tensor for the PE phase, computed using the GGA functional. We note that there are no significant differences between the BEC values obtained with the GGA and HSE functionals.

For consistency, we use the BEC values from the PE phase to calculate the polarization in both the FE and AFE configurations. 
The BEC tensor for the FE phase is reported in Section S4 the SM. As expected, the BEC components are positive for the cations and negative for the anions. The BEC values are large, deviating from the nominal ionic charges (namely, $+3$ for Sb, $-1$ for I, and $-2$ for S) indicating substantial charge redistribution across the bonds compared to an ideal ionic model, in particular along the $z$ direction. The highest BEC values are obtained for the Sb atoms, which play the crucial role of most ``active sites'' in the material's polarization, as discussed earlier. 

The polarization in the FE phase, calculated using the BEC method, is  $\simeq$ 30 $\mu$C/cm$^2$, a value very close to the one obtained through the Berry phase approach, confirming the reliability of the approximation. In the AFE phase, the polarization of a single chain is $\simeq$ 30 $\mu$C/cm$^2$, indicating that the FE and AFE phases exhibit similar polarization values along the $c$-axis, despite the overall macroscopic polarization of the latter phase being zero due to the antiparallel alignment of adjacent chains.

\noindent 
\begin{table*}[t!]
\begin{centering}
\begin{tabular}{|c|c|c|c|c|c|c|c|c|c|}
\hline 
 & $Z^{*}_{xx}$ & $Z^{*}_{xy}$ & $Z^{*}_{xz}$ & $Z^{*}_{yx}$ & $Z^{*}_{yy}$ & $Z^{*}_{yz}$ & $Z^{*}_{zx}$ & $Z^{*}_{zy}$ & $Z^{*}_{zz}$ \tabularnewline
\hline 
Sb  & 1.89  & 0.04  & 0  & 0.94  & 3.67  & 0  & 0  & 0  &   6.84\tabularnewline
\hline
S  & -1.09  &  -0.07  & 0  &  -0.88  & -1.76  & 0  & 0  & 0  &  -3.83\tabularnewline
\hline
I & -0.80  &  -0.43  & 0  & -0.81  &  -1.90  & 0  & 0 &  0 &  -3.01\tabularnewline
\hline
\end{tabular}
\par
\caption{BEC in the PE phase calculated within the GGA.}
\label{BEC_SbSI}
\end{centering}
\end{table*}


\begin{figure}[t!]
    \centering
    \begin{minipage}{0.8\textwidth} 
        \centering
        \begin{overpic}[width=0.55\textwidth,angle=270]{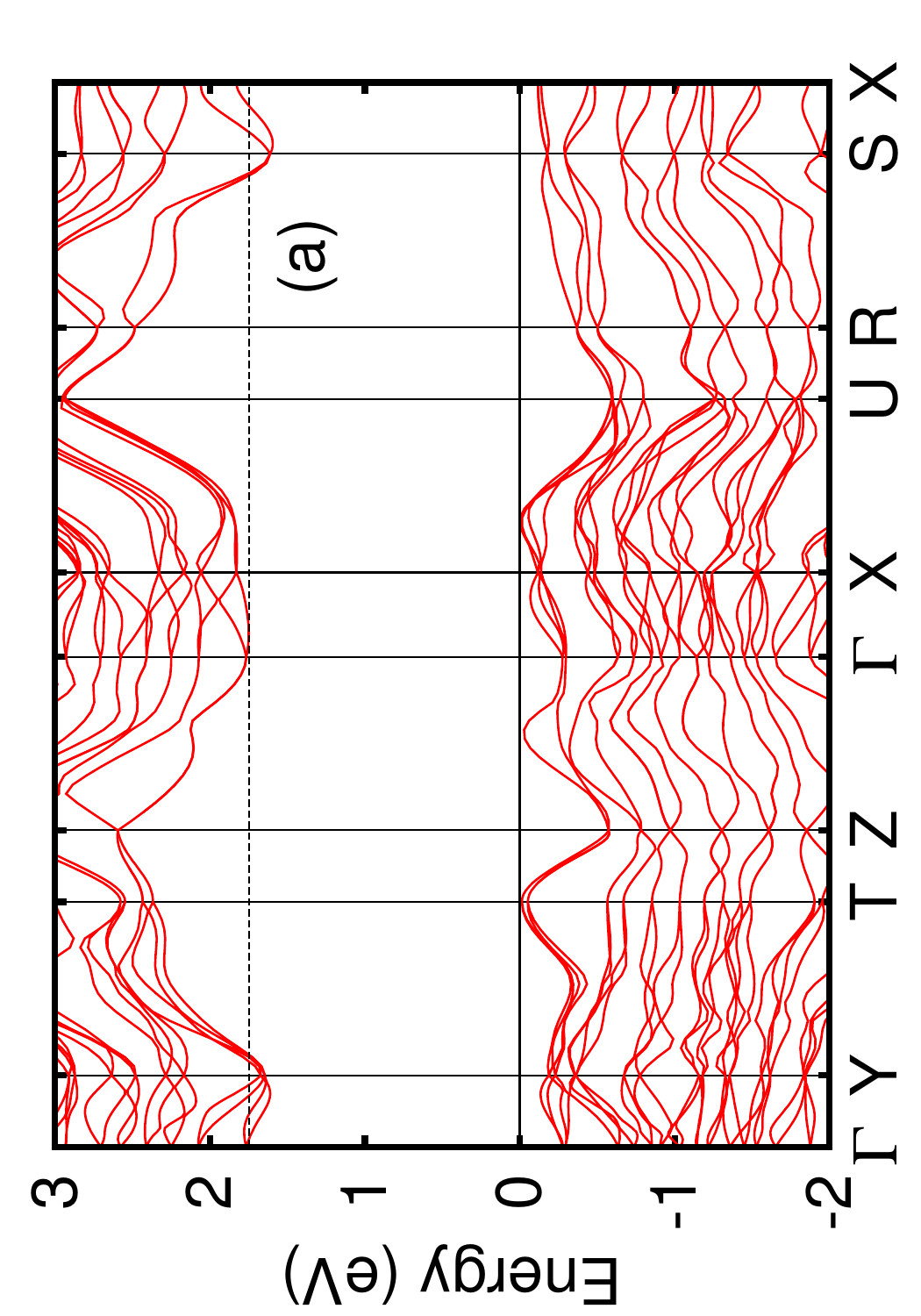}
            \put(97,5){ 
                \rotatebox{0}{ 
                    \includegraphics[width=3.5cm]{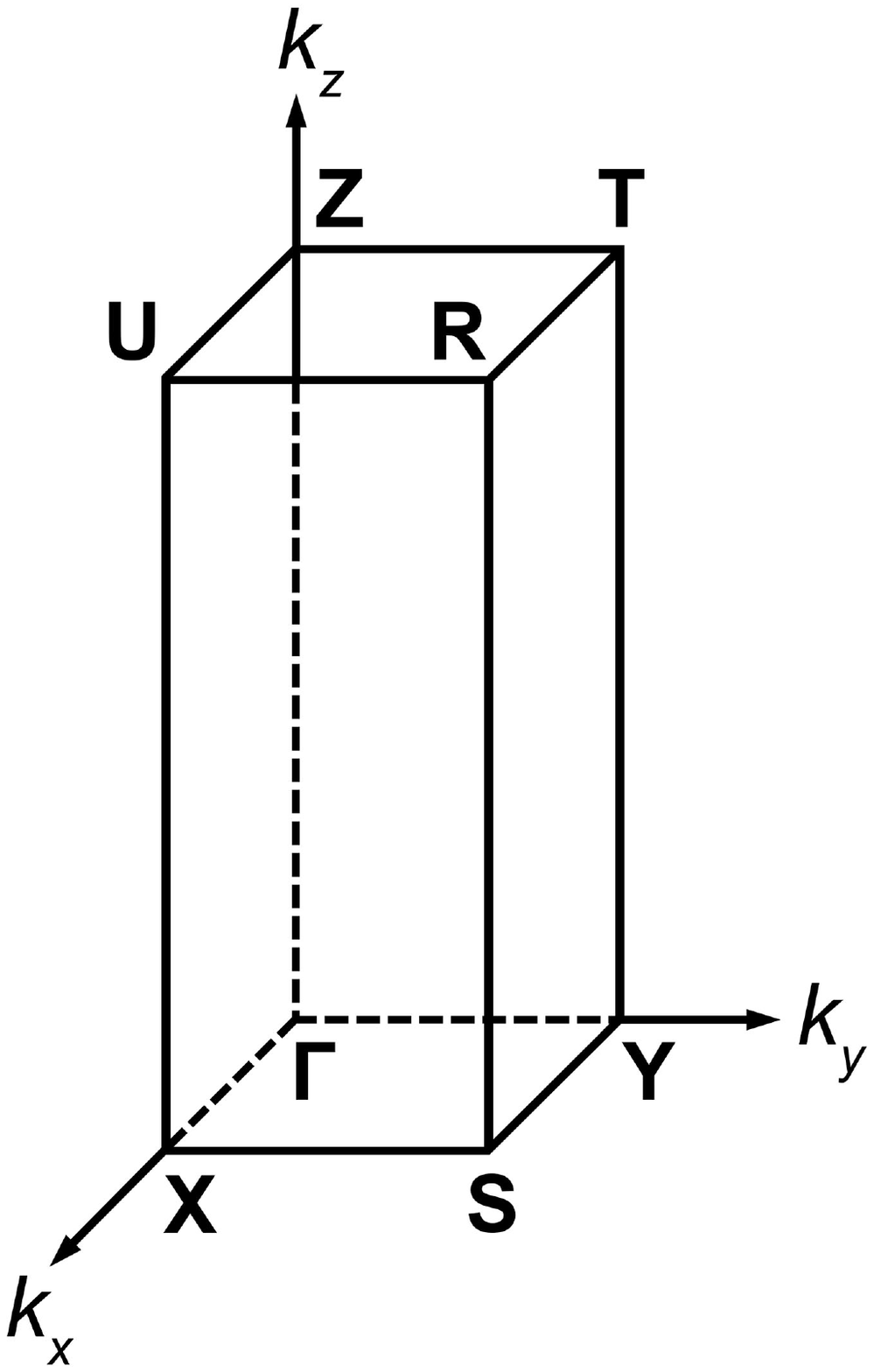} 
                }
            }
        \end{overpic}
        \caption*{} 
    \end{minipage}

    \vspace{1cm} 
    
    \begin{minipage}{0.8\textwidth} 
        \centering
        \includegraphics[width=0.55\textwidth,angle=270]{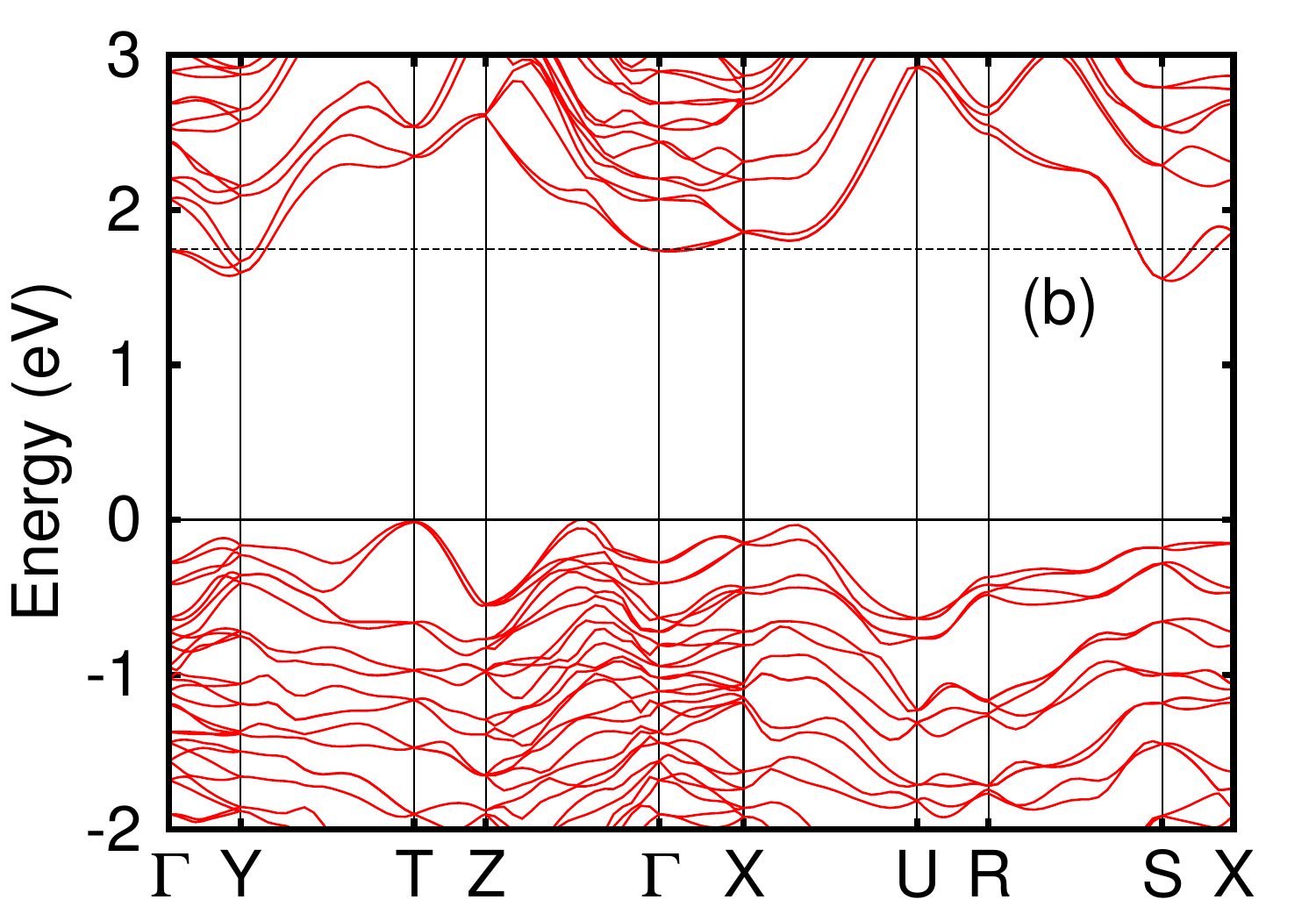}
        \caption*{} 
    \end{minipage}

    \caption{DFT band structure for the FE phase (a) and for the AFE phase  (b). The inset displays the Brillouin zone. The Fermi energy is set at 0 eV. The horizontal dashed line indicates the energy surface where the spin textures are calculated.}
    \label{Bands_SbSI}
\end{figure}

\subsection{Band structure}\label{sec.bands}
The band structures of the FE and AFE phases, calculated within the GGA and including SOC, are displayed in Fig. \ref{Bands_SbSI}(a) and (b), respectively. The results for the FE phase are consistent with previous studies \cite{Amoroso16,Butler15,Butler16}. The bands are quite wide along paths in the Brillouin zone parallel to the $k_z$ direction (e.g., Z-$\Gamma$ line or X-U), while they narrow along the in-plane directions (e.g., $\Gamma$-X line), reflecting the weak inter-chain hybridization. Compared to the GGA, HSE gives a rigid upward shift of the conduction bands without significantly modifying the main electronic features and the overall band dispersion. The HSE band structure is shown in Section S5 of the SM.  

The band gap in the FE phase is indirect [see Fig. \ref{Bands_SbSI}(a)]. The minimum of the conduction band is at the S point of the Brillouin zone, while three maxima of the valence band, with similar energies, are located at the T point and along the X-U and Z-$\Gamma$ line. The size of this indirect gap is 1.61 eV in the GGA, and increases to 2.29 eV when using HSE. A direct gap exist along the $\Gamma$-Y line and is equal to 1.75 eV in the GGA and 2.59 eV in HSE. All these gap values fall within the visible energy range of the electromagnetic spectrum, confirming that SbSI is potentially suitable for photovoltaic applications. 

Compared to the experimental value of 1.88 eV \cite{Madelung04}, the band gap of the FE phase is only slightly underestimated in the GGA, and conversely, drastically overestimated by HSE. However, it is important to note that the experimental band gap is determined by excitonic effects, which reduce the optical gap with respect to the single-particle one. These effects are not accounted for in the DFT band structure. Thus, the apparent agreement between the GGA result and experimental value may be due to a fortuitous compensation between the well-known band gap problem of DFT and the omission of exciton binding energies in the calculations. 

In the AFE phase [see Fig. \ref{Bands_SbSI}(b)], the size of the band gap, as well as the positions of the valence band maximum and conduction band minimum, remain similar to those in the FE phase. Hence, we expect light adsorption to occur in the same energy range, making the AFE phase equally suitable for exploiting the BPVE.

\begin{figure}
\includegraphics[width=8.5cm,angle=0]{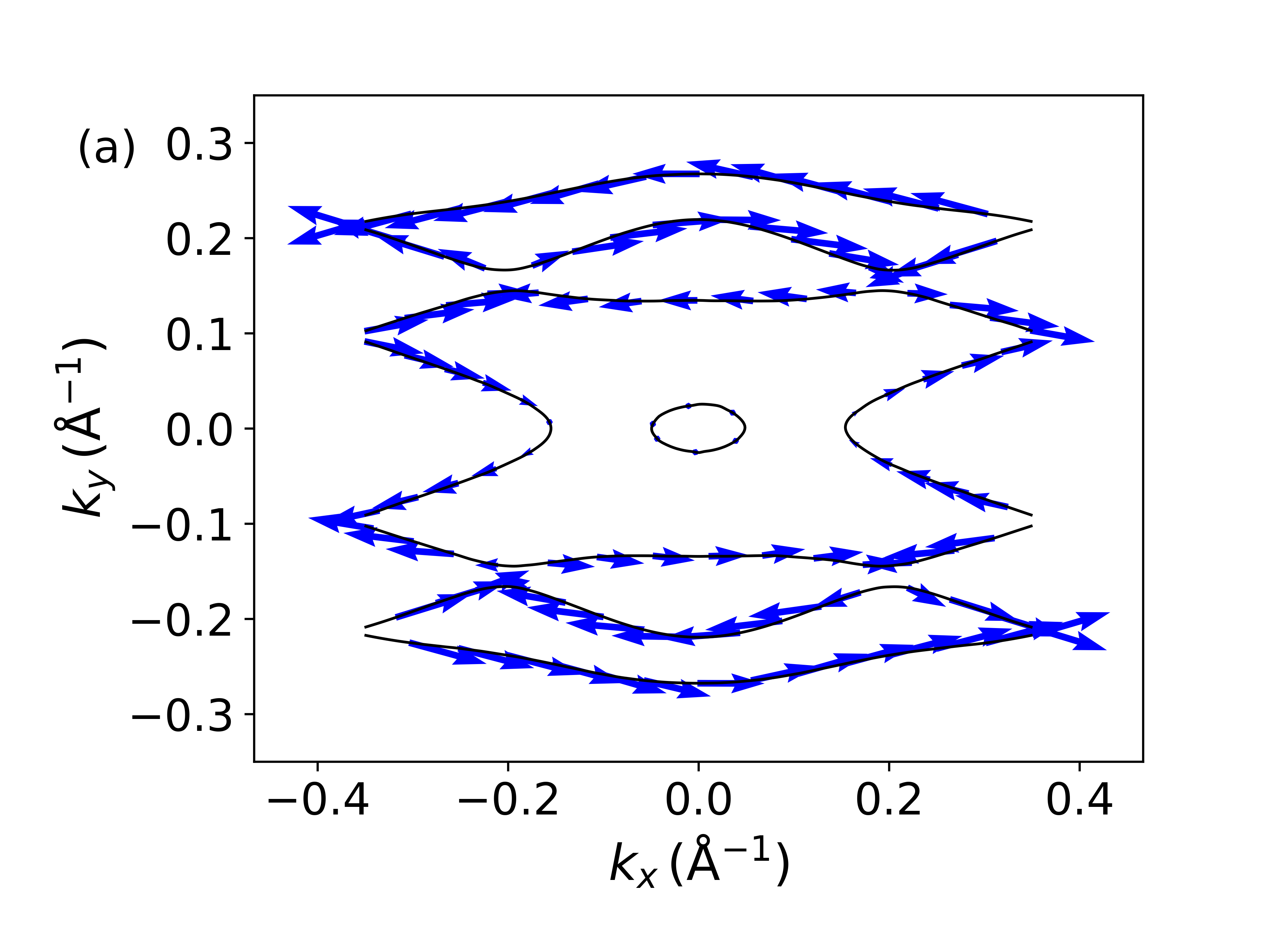}
\includegraphics[width=8.5cm,angle=0]{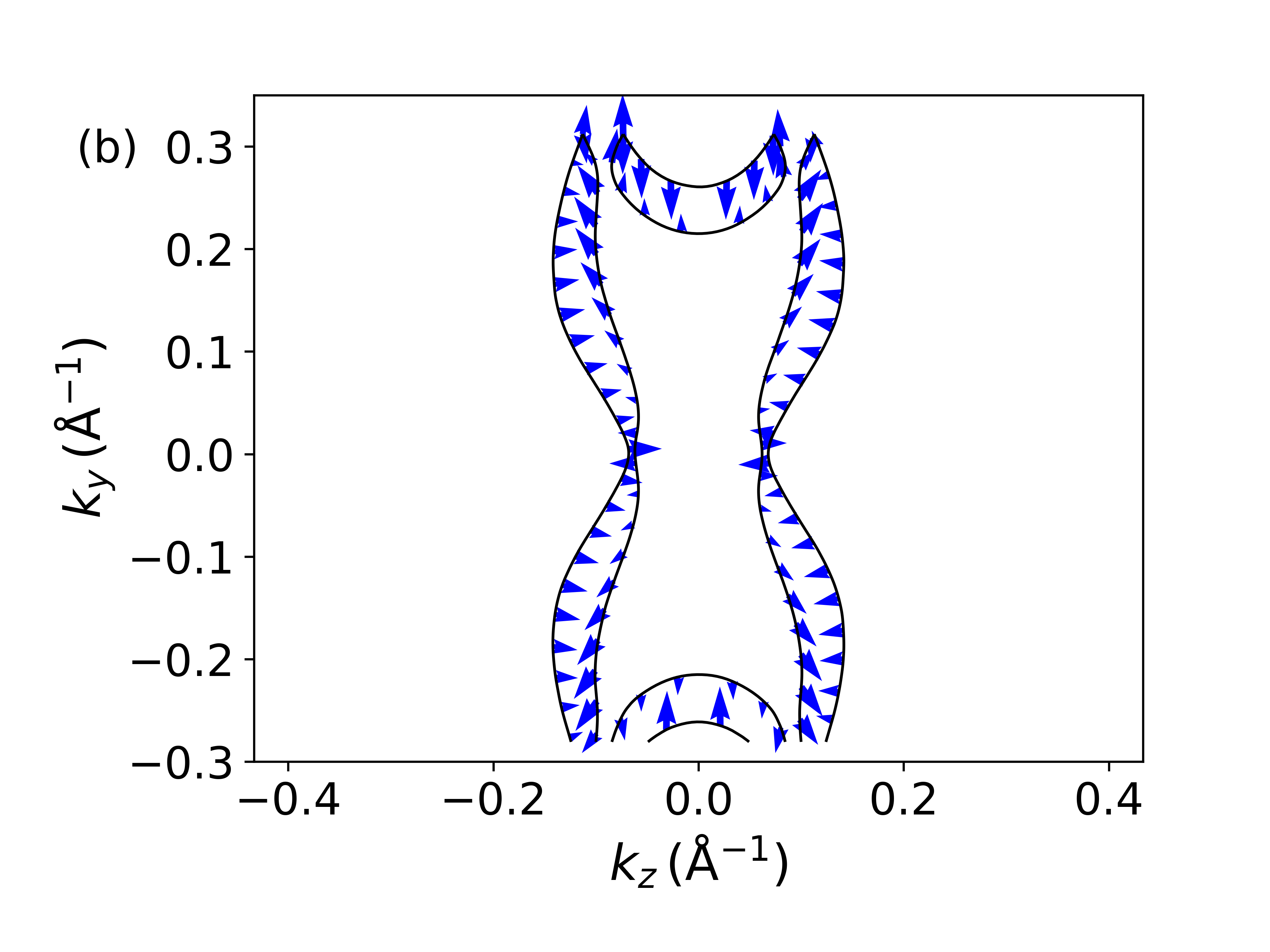}
\includegraphics[width=8.5cm,angle=0]{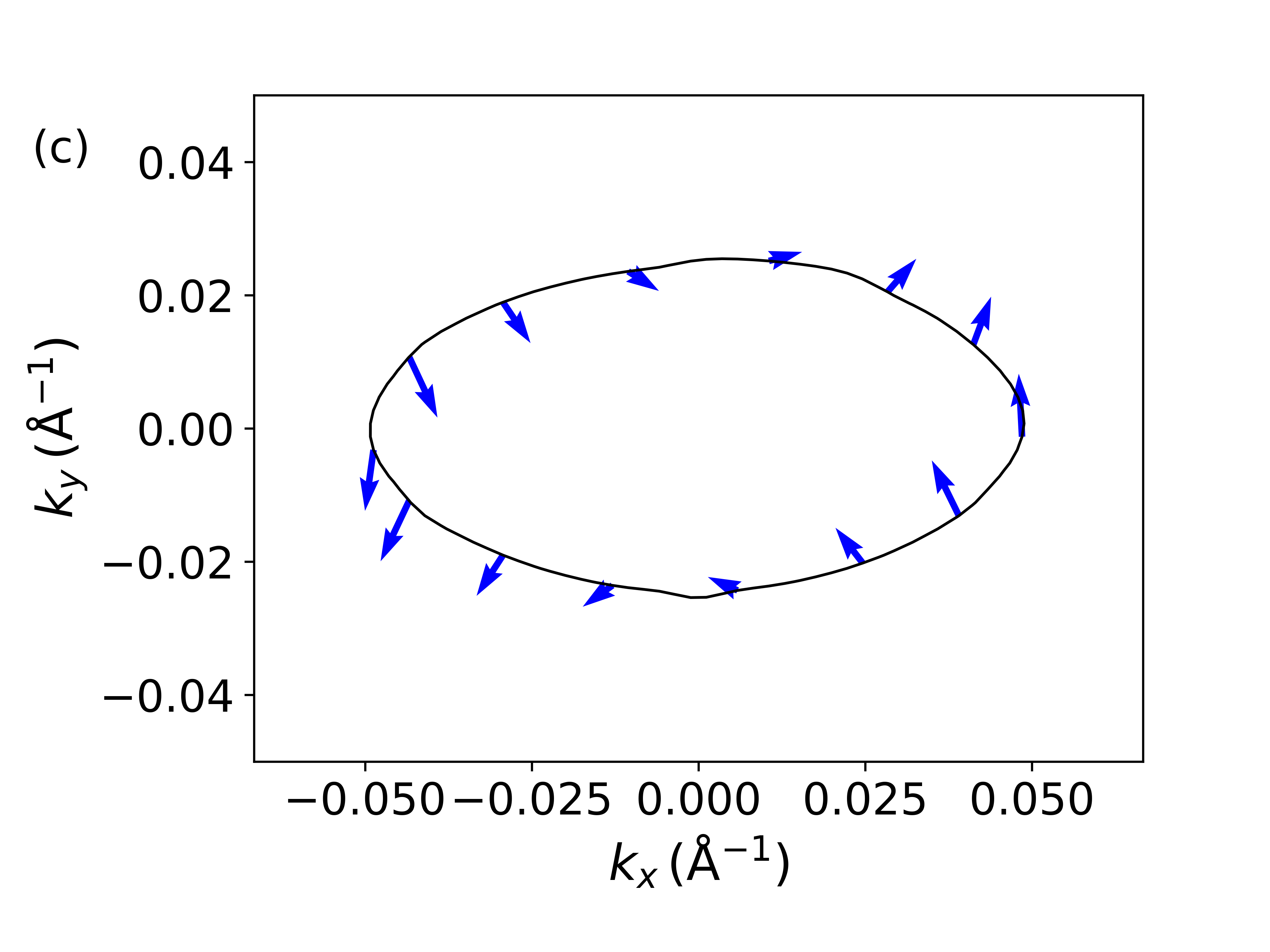}
\includegraphics[width=8.5cm,angle=0]{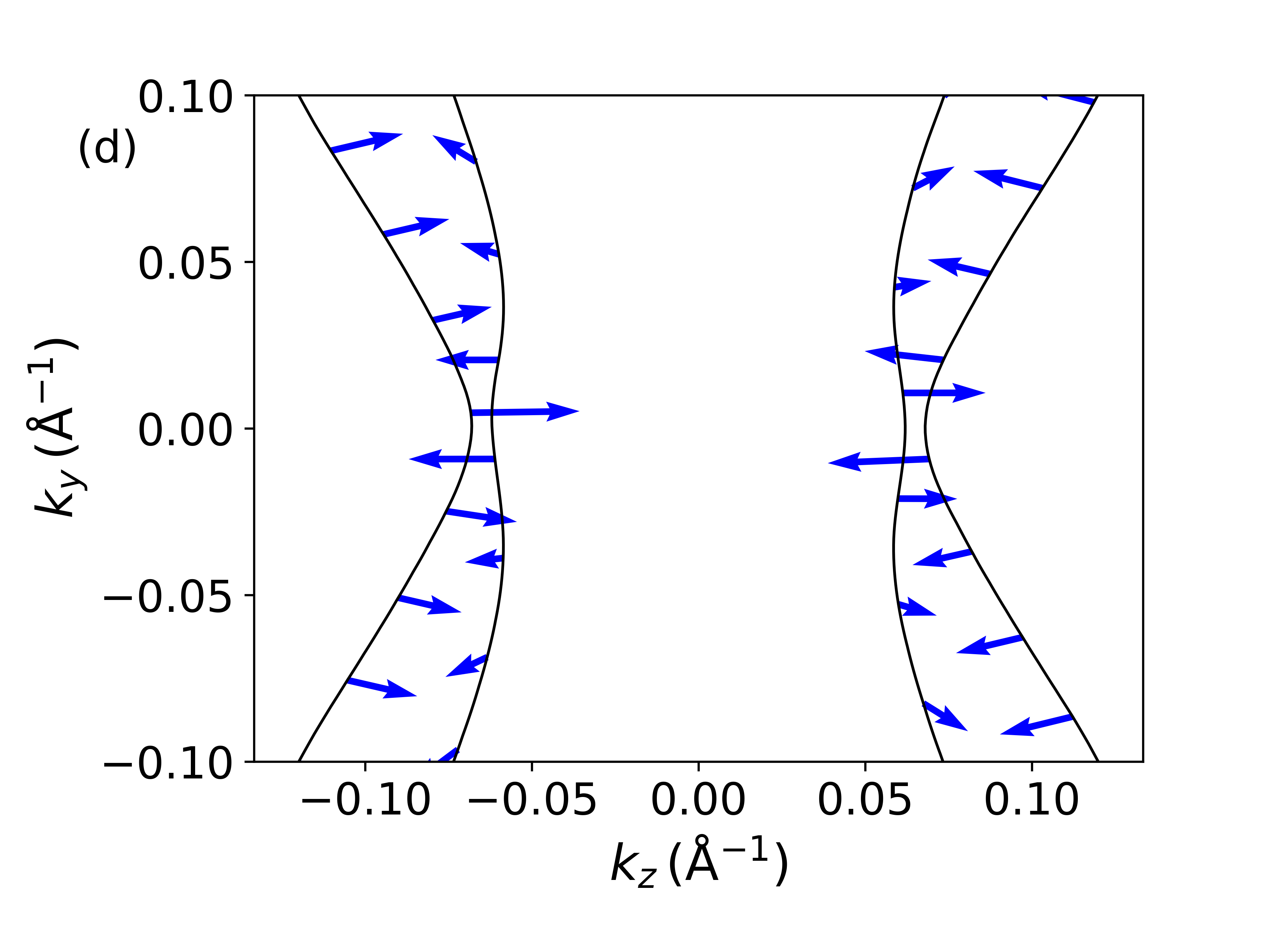}
\caption{Spin textures calculated by DFT. (a) plot in the $k_z=0$ plane at the constant energy surface $E-E_\mathrm{VBM}$=1.75 eV above the valence band maximum for the FE phase. (b) plot in the $k_x=0$ plane at the constant energy surface $E-E_\mathrm{VBM}$=1.75 eV for the AFE phase. (c) and (d) are zoomed-in views of panels (a) and (b) around the $\Gamma$ point.} \label{spin_textures}
\end{figure}

The main differences between the FE and AFE phases are observed in some bands that are degenerate in one phase but split by the SOC in the other. This is a direct consequence of the change in the crystal space group during the phase transition. A particularly insightful analysis can be made around the $\Gamma$ point; despite the intricate band structure, one can resort to an effective linear spin-momentum Hamiltonian for a qualitative understanding. In the FE phase, the point group symmetry at $\Gamma$ is $C_{2v}$. Thus, the effective SOC terms acquire the form $\alpha_{xy} k_x\tau_y$ and $\alpha_{yx} k_y\tau_x$, where $\tau_x$ and $\tau_y$ are the Pauli spin matrices, and $\alpha_{xy}$ and $\alpha_{yx}$ are coupling parameters representing the non-zero components of a rank-two spin-momentum coupling pseudotensor \cite{Ganichev14}. These effective SOC terms can be-rewritten in the standard Rashba and Dresselhaus forms by performing an appropriate unitary transformation and lead to a characteristic in-plane spin texture \cite{Amoroso16}. Because of this peculiar feature, SbSI is referred to as a FE Rashba-Dresselhaus semiconductor. 

In contrast, in the AFE phase, the point group symmetry at $\Gamma$ is $D_{2}$, leading to a different effective Hamiltonian, corresponding to a spin texture with spin components that can also point outside the plane, i.e. along the $z$ axis. The effective SOC terms now take the form $\alpha_{xx} k_x\tau_x$, $\alpha_{yy} k_y\tau_y$, and $\alpha_{zz} k_z\tau_z$, where $\alpha_{xx}$, $\alpha_{yy}$, $\alpha_{zz}$ are the non-zero components of a spin-momentum coupling pseudotensor consistent with the $D_{2}$ symmetry. These SOC terms give rise to a Dresselhaus-Weyl spin texture \cite{Acosta}. The effect of the $\alpha_{zz} k_z\tau_z$ term is particularly evident in the band structure, where bands along the Z-$\Gamma$ direction, which are two-fold degenerate in the FE phase, become spin-split in the AFE phase. 

The spin textures obtained from the DFT calculations are shown in Fig. \ref{spin_textures}, with the blue arrows representing the directions of the expectation values of the Pauli spin matrices. In particular, panel (a) shows the plot in the $k_z = 0$ plane at the constant energy surface $E - E_\mathrm{VBM} = 1.75$ eV above the valence band maximum (VBM) (see the horizontal dashed line in Fig. \ref{Bands_SbSI}) for the FE phase, while panel (b) shows the plot in the $k_x = 0$ plane at the same constant energy surface for the AFE phase. Panels (c) and (d) are zoomed-in views of panels (a) and (b) around the $\Gamma$ point. In panel (c), we clearly observe the characteristic Rashba-Dresselhaus spin texture, consistent with previous reports for the FE phase \cite{Amoroso16}. In contrast, panel (d) shows, in the AFM phase, the emergence of a spin-z polarization, which is absent in the FE phase.  

The observed differences in the band structure and, even more, in the spin texture between the two phases reveal a particularly interesting effect. Since the FE and AFE phases are energetically competitive, it might be possible to explore means to stabilize one phase over the other, thereby tuning the effective spin-charge coupling and, consequently, the spintronic properties of the material. In the AFE phase, in particular, the spin texture with $\sim k_z\tau_z$ character is expected to give rise to the so-called ``unconventional'' current-induced spin polarization \cite{Calavalle2022, Roy2022, Gupta2024}, where a charge current induces spin polarization of the conduction electrons along the current direction. This phenomenon, recently reported in Te, has drawn significant attention for the design of spintronic devices \cite{Calavalle2022}. In SbSI, it can be further combined with the BPVE, 
investigated in the following, resulting in the integration of multiple functionalities.

\subsection{Bulk photovoltaic effect}\label{sec:BPVE}

A photocurrent density $\mathbf{j}$ driven by an electric field $\mathbf{E}(\omega)$ of frequency $\omega$ in the non-magnetic material SbSI can be written as \cite{Wang20,Puente23}

\begin{align}
j^a &= 2 \sum_{bc} \sigma^{abc}(\omega) \, \mathrm{Re}[E^b(\omega) E^c(-\omega)] \nonumber \\
    &\quad + t \sum_{bc} \eta^{abc}(\omega) \, \mathrm{Im}[E^b(\omega) E^c(-\omega)] \label{eq:current}
\end{align}
where the first term represents the shift current (linear photogalvanic effect) and the second term corresponds to the injection current (circular photogalvanic effect). Both currents are expected to appear in both the FE and AFE phases, due to their non-centrosymmetric space groups, and more specifically, to their gyrotropy. $\sigma^{abc}(\omega)$ denotes a component of the shift photoconductivity tensor, which is real, while $\eta^{abc}(\omega)$ denotes a component of the circular injection photoconductivity tensor, which is imaginary. They are defined in Section S7 of the SM following Refs. [\cite{Azpiroz18}] and [\cite{Puente23}].
The indices $a,b,$ and $c$ label vector and tensor components; in particular, $a$ indicates the direction of the photocurrent. $t$ represents the electron relaxation time. The shift current corresponds to the dc part of the second-order response to the dipole electric field of light. In contrast, the injection photoconductivity corresponds to the rate of change of the interband dc part of the second-order response.

The photoconductivity tensors presented in the following are all computed starting from a Wannier interpolation of the GGA band structure calculated with SOC, as detailed in Section \ref{sec.comp_details}. 
The results obtained from the HSE band structure are very similar to the GGA ones, except for the position of the photoconductivity onset energy, which is shifted due to the (roughly rigid) upward shift of the conduction bands and related increase of the band gap. Due to this uncertainty on the band gap size, in the following, we will present all the results with the onset energy aligned with the experimental band gap value, as for common practice in the literature \cite{Azpiroz18}. Since the value of the band gap in the AFE phase is similar to that in the FE phase, we shift the photocurrents in AFE configuration by the same amount.

\begin{figure}[t!]
 \hspace*{-1cm}
    \centering
    \begin{minipage}{0.4\textwidth} 
        \begin{overpic}[width=0.87\textwidth,angle=270]{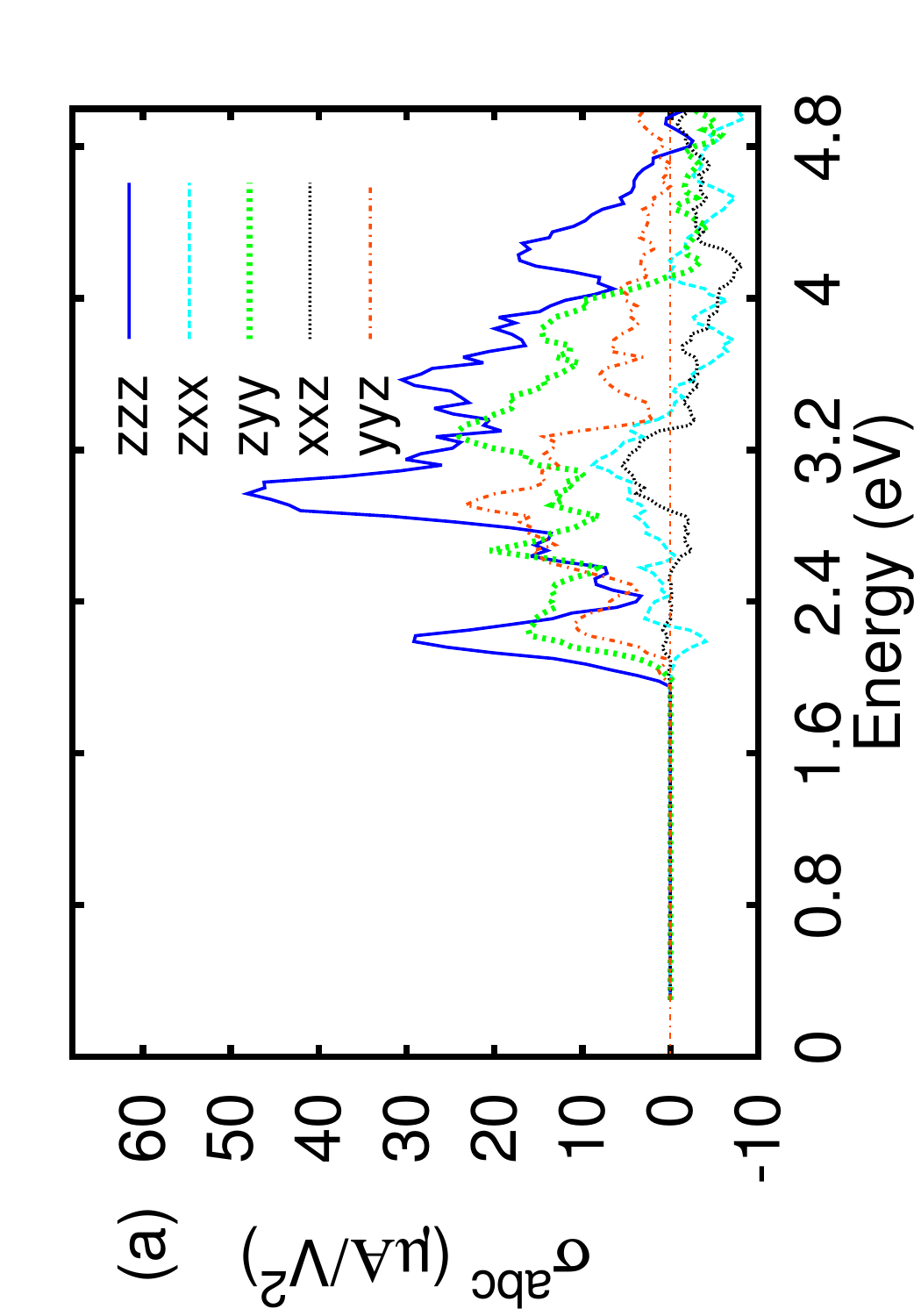}
            \put(19,66){ 
                \rotatebox{270}{ 
                    \includegraphics[width=2cm]{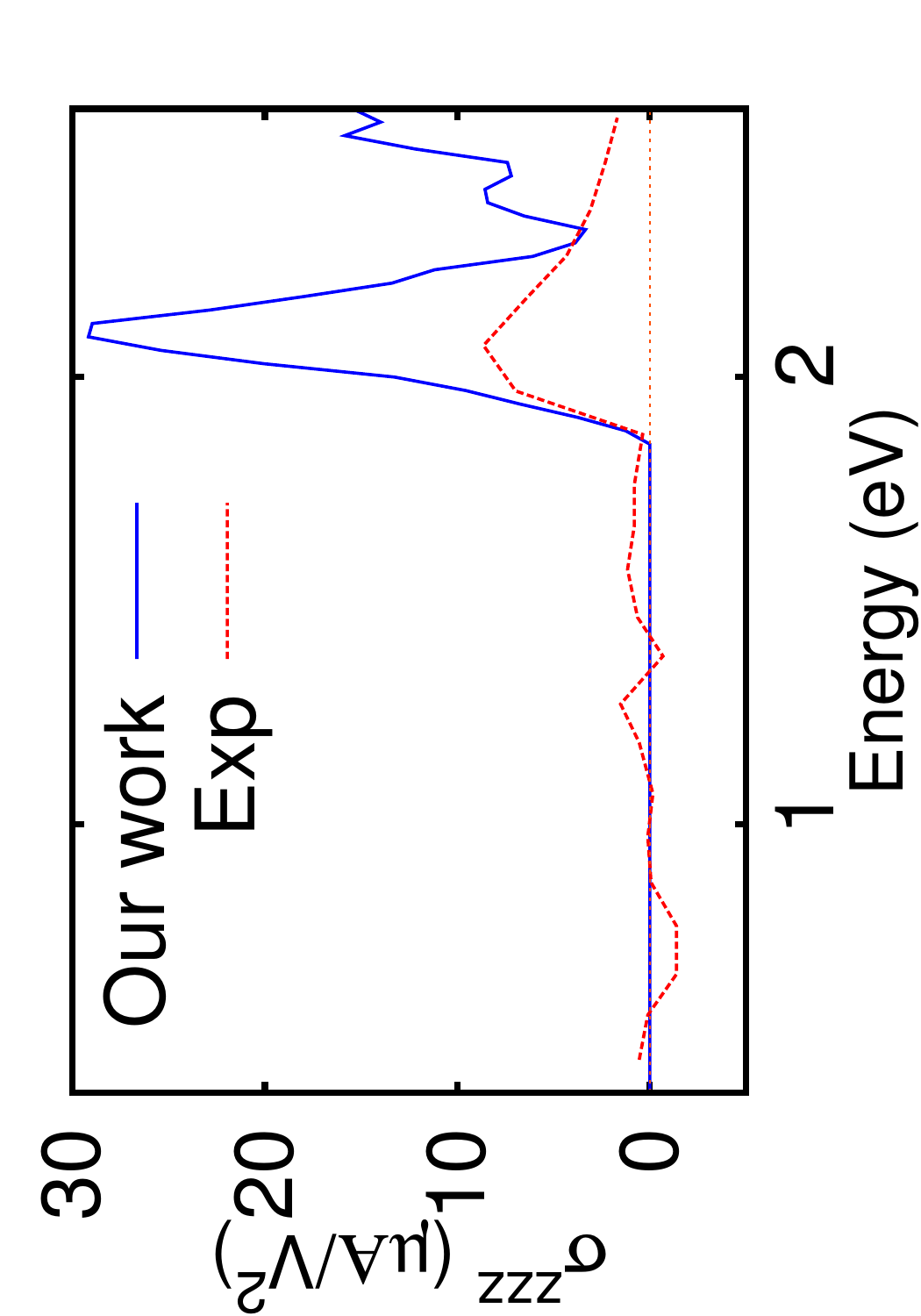} 
                }
            }
        \end{overpic}
        \caption*{} 
    \end{minipage}
    \hspace{0.06\textwidth} 
    \begin{minipage}{0.4\textwidth} 
        \centering
        \includegraphics[width=0.87\textwidth,angle=270]{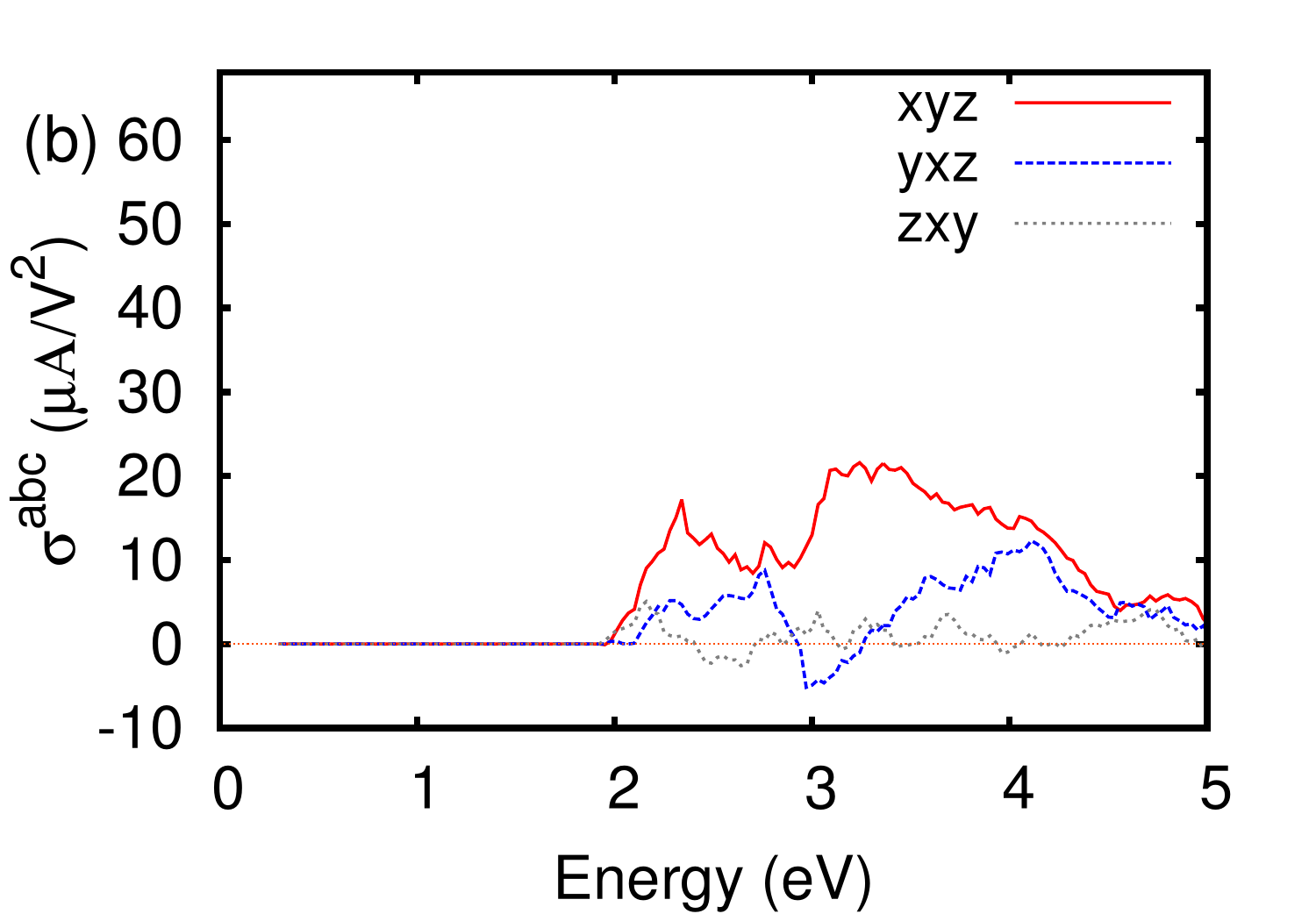}
        \caption*{} 
    \end{minipage}
    
    \caption{Non-zero independent components of shift photoconductivity tensor for the FE phase (a) and the AFE phase (b). The inset shows a comparison between our result for the $\sigma^{zzz}$ and the experimental data from Ref. \cite{Sotome19}.}
    \label{SHC_Ferro_Antiferro}
\end{figure}

\subsubsection{Linear photovoltaic effect}
In the FE phase, the shift photoconductivity tensor of SbSI has five
independent non-zero components, namely $\sigma^{zzz}$, $\sigma^{zxx}$, $\sigma^{zyy}$, $\sigma^{xxz}$ and $\sigma^{yyz}$, as allowed by the system's space group, Pna2$_1$. They are displayed in Fig. \ref{SHC_Ferro_Antiferro}(a) as a function of the energy, $E=\hbar\omega$.
All other components can be obtained from these independent components by exchanging the last two indices, related to the light polarization.
The largest component is $\sigma^{zzz}$, corresponding to the electric field being polarized along the $c$ axis and the current flowing in the same direction. The shift current is thus larger when the incident light is parallel to the FE distortion direction compared to the case when it is perpendicular. In the visible energy region ($E\sim 1.6-3.3$ eV), $\sigma^{zzz}$ displays two prominent peaks, reaching about 30 $\mu$A/V$^2$ at 2 eV and 47 $\mu$A/V$^2$ at 3 eV. These peak values are significantly higher than those for prototypical FE compounds, like for BiFeO$_3$ 
and BaTiO$_3$, where the maximum photoconductivities in the visible energy range are 0.05 $\mu$A/V$^2$ and 5 $\mu$A/V$^2$, respectively \cite{Young12}. Hence, SbSI is potentially an excellent material candidate for bulk photovoltaic conversion. 

The calculated $\sigma^{zzz}$ can be compared to the experimental data \cite{Sotome19}, as shown in the inset of Fig. \ref{SHC_Ferro_Antiferro}(a). While the shape of the first peak is qualitatively well reproduced, the experimentally measured maximum is significantly lower than the predicted one. Such discrepancies between measured and calculated shift photoconductivities have been reported in other studies on different materials. They are often attributed to the presence of multiple ferroelectric domains with opposite spontaneous polarizations in the sample, which reduces the overall shift current by averaging out the contributions from individual domains \cite{Chang23}.

When transitioning to the AFE phase, the shift photoconductivity tensor is found to change compared to the FE case acquiring different non-zero components, as a result of the change in the crystal space group (see SM). 
Specifically, the independent components become $\sigma^{xyz}$, $\sigma^{yxz}$ and $\sigma^{zxy}$. They are displayed in Fig. \ref{SHC_Ferro_Antiferro}(b). 
The largest one is $\sigma^{xyz}$. Thus, the shift current in the AFE phase is expected to be maximized in the $x$ direction, which is perpendicular to the chains, conversely to what is predicted in the FE phase.
$\sigma^{xyz}$ presents peaks of approximately 17 $\mu$A/V$^2$ near 2 eV and 20 $\mu$A/V$^2$ at 3 eV. These photoconductivity values are lower than those in the FE phase at similar energies, but still large if compared to the response of the prototypical bulk FE oxides \cite{Young12}. 

To our knowledge, the shift current in the AFE phase has not been measured. Therefore, we lack benchmark experimental results for comparison. However, since the AFE phase exists near room temperature, it should be experimentally accessible. Moreover, given that the non-zero components of the shift photoconductivity tensor differ between the FE and AFE phases, we suggest that measuring the shift conductivity could serve as a method to determine the phase of the material and follow the phase transition.

\subsubsection{One-chain system}

The maximum shift photoconductivity can be further enhanced by moving from bulk 3D materials to low-dimensional systems. For example, very large photoconductivities were predicted in two-dimensional ferroelectrics \cite{Rangel17}, like monolayer GeS and GeSe, and were related to the presence of van Hove singularities in the materials density of states (DOS) \cite{Cook17,Rangel17}. In the case of SbSI, we can explore a similar idea by considering the extreme limit of one isolated chain, thus reducing the material to 1D. 

The one-chain system is modelled considering the FE unit cell, removing one of the two chains, and increasing the lattice parameters along the transverse directions to $a=b=12$ {\AA}, to avoid interaction between periodically repeated images. Then, following Refs. [\cite{Azpiroz18}] and [\cite{Rangel17}], the shift photoconductivity tensor components for the single chain $\sigma^{abc}_{SC}$ are estimated by rescaling the results from the 3D supercell as $\sigma^{abc}_{SC} = \frac{a}{w_{x}} \frac{b}{w_{y}} \sigma^{abc}$ where $w_{a}$ and $w_{b}$ are the widths of the chain along the $x$ and $y$ directions, taken approximately equal to 2.62 {\AA} and 6.57 {\AA} (see SM). 
 
\begin{figure}[t!]
\centering
\includegraphics[width=6.5cm,angle=270]{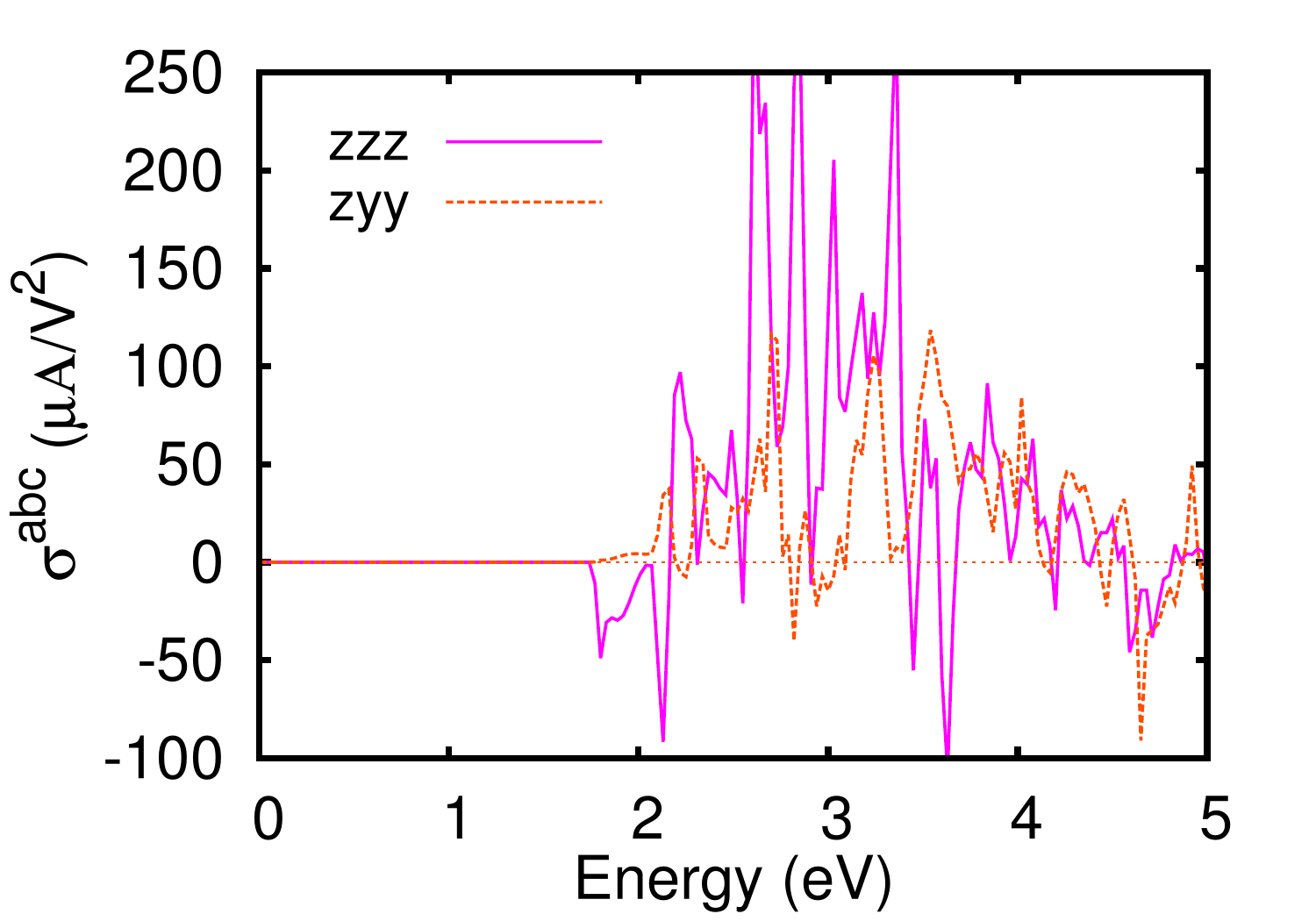}
\caption{Non-zero components of the shift photoconductivity tensor for the one-chain system.} \label{SHC_SUC}
\end{figure}

The results for the one-chain system are displayed in Fig. \ref{SHC_SUC}, where only the components $\sigma^{zzz}$ and $\sigma^{zyy}$ are non-zero. In particular, $\sigma^{zzz}$ reaches peak values one order of magnitude larger than those for the bulk systems, especially in the energy range between 2.5 and 3.2 eV. The shape of the spectrum is characterized by sharp peaks which can be clearly related to the van Hove singularities in the DOS, which is shown in Section S6 of the SM. 
This results confirms the significant role these singularities play in enhancing the shift photoconductivity, similar to what has been reported in monolayer ferroelectrics. 

Although isolating a single chain in experiments is challenging, it may be possible to fabricate nanowires with diameters of several nanometers, consisting of a few chains. This approach has already been demonstrated in tellurium, a material similarly composed of weakly van der Waals-bonded chains \cite{Calavalle2022}.  Nanowires of SbSI could exhibit photoconductivities that are intermediate between those of bulk and single-chain systems, thus offering exceptionally high photovoltaic performance.

\subsubsection{Circular photogalvanic effect}
 
\begin{figure}[t!]
\centering
\includegraphics[width=5.5cm,angle=270]{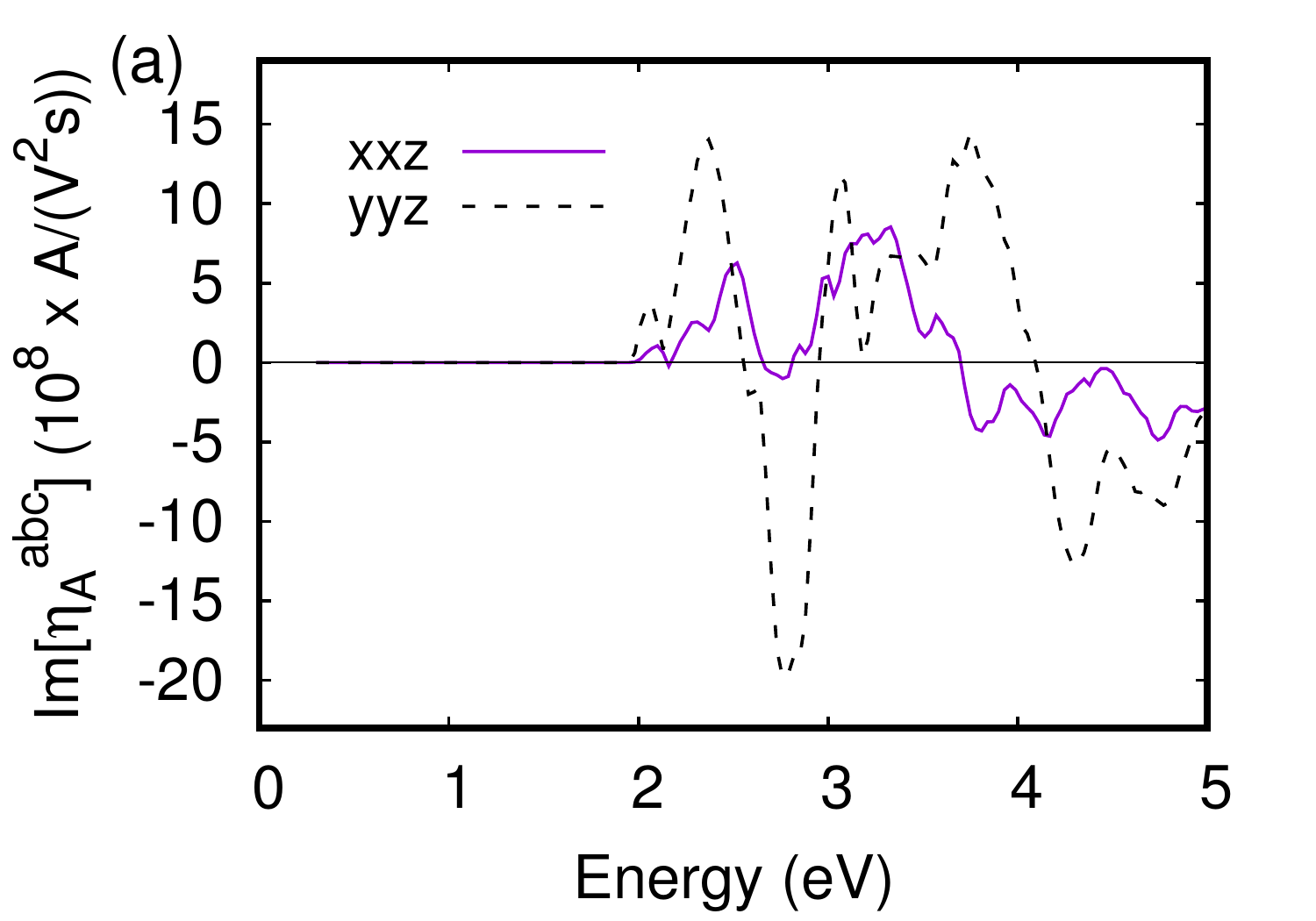}
\includegraphics[width=5.5cm,angle=270]{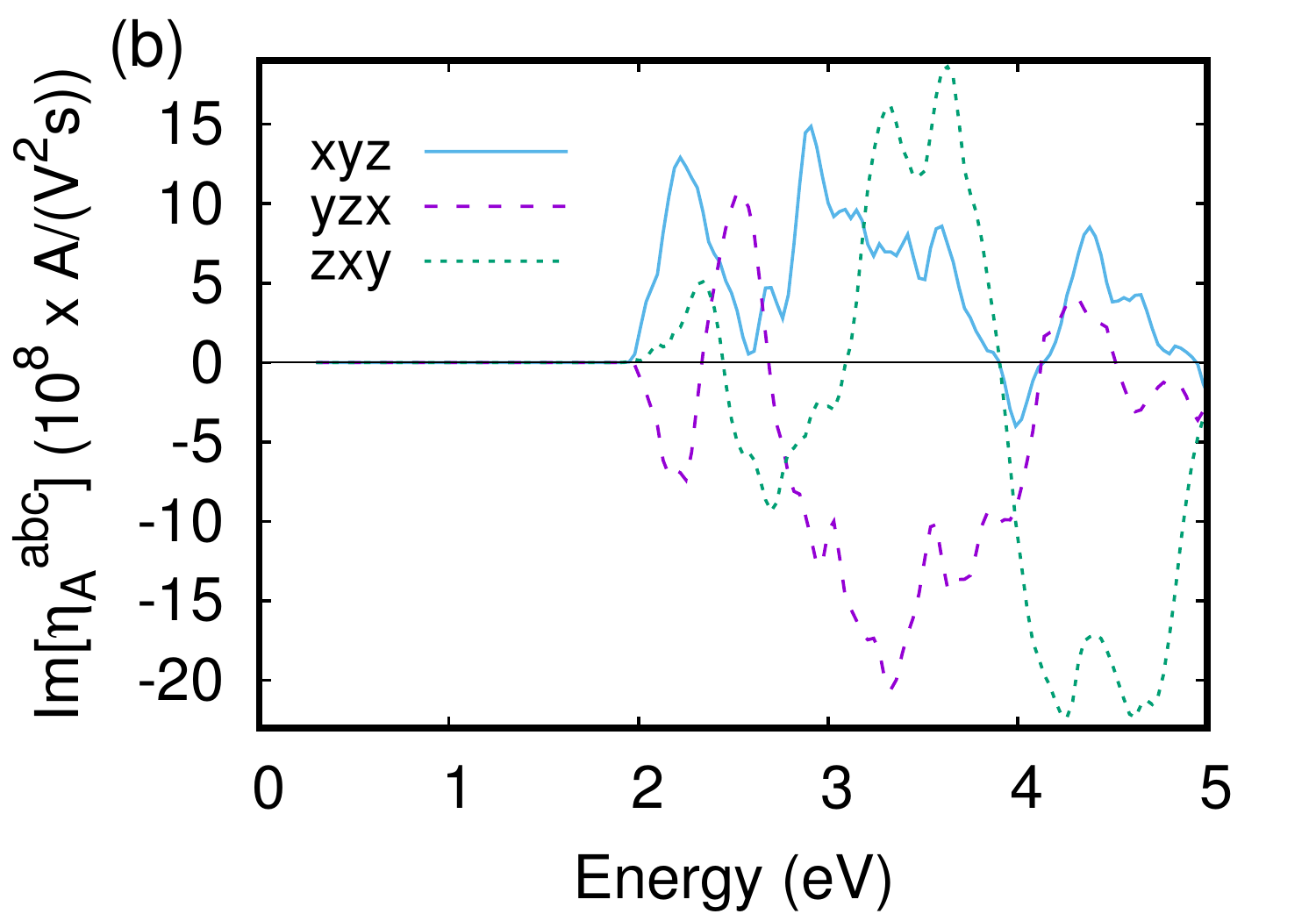}
\caption{Non-zero independent components of the circular injection photoconductivity tensor for (a) the FE phase, and (b) the AFE phase.} \label{CIC_FE_AFE}
\end{figure}

In the FE state, the circular injection photoconductivity tensor has two independent components, namely $\eta^{xxz}$ and $\eta^{yyz}$, allowed by symmetry, as shown in Fig. \ref{CIC_FE_AFE}(a). All other non-zero components are obtained from these by exchanging the last two indices and reversing the sign (see Section S7-C-2 in the SM). Among these two independent components, $\eta^{yyz}$ is considerably larger than $\eta^{xxz}$ in the visible light energy range. It presents a first small peak of approximately 4 $\times$ 10$^8$ A/V$^2$s at around 2.1 eV, followed by a significantly larger peak of 14 $\times$ 10$^8$ A/V$^2$s at around 2.4 eV, and a sharp negative peak reaching -19 $\times$ 10$^8$ A/V$^2$s at about 2.8 eV. As the energy increases, further large peaks are observed with the absolute value of the photoconductivity remaining quite large even outside the visible range.

It is particularly important to note that all $\eta^{zbc}$ components in bulk 
FE SbSI are zero (for any $a$ and $b$ indices), indicating that the circular injection current is absent along the chains, but can only emerge in the transverse $x$ and $y$ directions, regardless of the light polarization. Interestingly, although the electronic bands are relatively flat in these perpendicular directions (i.e., the electron group velocity is quite small) owing to the weak inter-chain hybridization, the obtained photoconductivity peak values are an order of magnitude higher than those reported in other bulk materials, such as CdSe, CdS \cite{Nastos10,Laman99,Laman05} and TaIrTe$_4$ \cite{Puente23}, which have been recently studied by DFT calculations. This indicates that simple arguments based on qualitative inspections of the band structure do not always hold in the analysis of photocurrents. Accurate quantum mechanical calculations are generally required to obtain reliable predictions. 

Turning to the AFE phase, we find that the independent components of the circular injection photoconductivity become $\eta^{xyz}$, $\eta^{yxz}$ and $\eta^{zxy}$, as shown in Fig. \ref{CIC_FE_AFE}(b). Thus, at variance with what happens in the FE phase, the circular injection current can be measured along all directions, including along the chains (longitudinal direction), for an opportune choice of the light polarization. 

The important result that the circular photogalvanic effect generates only transverse photocurrents in the FE phase, while both transverse and longitudinal photocurrents appear in the AFE phase, can be further confirmed through an alternative analysis. 
Specifically, the rank-three circular injection photoconductivity tensor can be recast into a rank-two pseudotensor of components $\gamma_{ab}\propto-i\epsilon_{bcd}\eta^{acd}$ (see Ref. \cite{Tsirkin}).
This so-called circular photogalvanic pseudotensor has the same structure as the optical activity tensor, and it transforms like any other rank-two pseudotensor \cite{Ganichev14,Droghetti22}, including the effective SOC pseudotensor introduced in Section \ref{sec.bands}, and whose components are the spin-momentum coupling parameters. Therefore, the form of the photogalvanic pseudotensor must mirror that of the spin-momentum coupling and, in turn, reflects the underlying spin texture.
In practice, this means that we can predict the nature of the circular photogalvanic effect by analyzing the spin texture — and vice versa \cite{Ganichev14}.

In the case of FE SbSI, we find that the only two non-zero components of the circular photogalvanic pseudotensor are $\gamma_{xy}$ and $\gamma_{yx}$, which are consistent with the non-zero spin–momentum coupling parameters $\alpha_{xy}$ and $\alpha_{yx}$, describing the Rashba–Dresselhaus in-plane spin texture. In contrast, in the AFE phase, there are three non-zero and independent components: $\gamma_{xx}$, $\gamma_{yy}$, and $\gamma_{zz}$, the latter accounting for the longitudinal photocurrent. These components correspond to the effective spin–momentum coupling parameters $\alpha_{xx}$, $\alpha_{yy}$, and $\alpha_{zz}$ identified in the band structure analysis, and are consistent with the emergence of spin-$z$ polarization alongside the in-plane spin texture. 

In summary, our results show that the circular photovoltaic effect can be used to probe the FE-AFE transition in a similar way to the linear photovoltaic effect, as the non-zero components of the photoconductivity change with the symmetry of the crystal structure. However, measuring circular injection currents can provide additional insights into the spin-dependent properties of the material. This feature could eventually be important for integrating the BPVE with spintronic functionalities.




\section{Conclusions}\label{sec:conclusions}

In this paper, we presented a comprehensive first-principles investigation of the structural, electronic, and photoconductive properties of SbSI, focusing in particular on both the linear and circular photogalvanic contributions to the BPVE.
Our analysis compares in detail the FE and the AFE phases, as the phase transition occurs near room temperature, making both phases experimentally accessible. 

We reproduced the crystal structures of the AFE and FE phases, elucidating the mechanisms responsible for their stabilization. 
In the FE phase, the atoms of the two chains of the unit cell shift in the same direction along the $c$-axis, resulting in a net polarization due to the shift of the cation sub-lattice with respect to the anion one. In contrast, in the AFE phase, the cations in adjacent chains move in opposite directions along $c$, leading to a zero net polarization.

The phase transition is accompanied by a modification in the band structure. While the size of the band gap and the positions of the valence band maximum and conduction band minimum in the Brillouin zone remain essentially  unchanged, the spin-texture transforms from Rashba-Dresselhaus in the FE phase to Dresselhaus-Weyl in the AFE phase. Notably, the latter case is expected to give rise to ``unconventional'' current-induced spin polarization, a phenomenon of interest for the design of spintronic devices.

To explore the BPVE, we performed a comprehensive analysis of both the shift and circular injection photoconductivity tensors, identifying all symmetry-allowed components and accurately estimating their magnitudes. The results can serve as a practical guide for distinguishing between the different phases of SbSI through photocurrent measurements. Moreover, we predict that the shift photoconductivity of SbSI in both phases can reach peak values significantly larger than those of most prototypical FE compounds, far exceeding those reported in experiments to date. This suggests the potential to improve the material's photovoltaic performance beyond the current state of the art.

Finally, we demonstrated that circular injection currents can serve as sensitive probes of the changes in spin texture, revealing a strong coupling between structural, optical, and spintronic properties.
 This finding underscores the promise of SbSI as a platform for exploring fundamental aspects of multifunctional behavior. 
 


\medskip
\textbf{Acknowledgements} \par The authors acknowledge Julen Iba\~nez Azpiroz and Alvaro Ruiz Puente for their technical support in the calculation of the photoconductivities.
This work was supported by the Next-Generation-EU program via the PRIN-2022 SORBET 1133 (Grant No. 2022ZY8HJY), the ICSC initiative (National Center for High-Performance Supercomputing, 1134 Big Data and Quantum Computing), the PE4 Partenariato Quantum Science and Technology (NQSTI).
High-performance computing resources and support were provided by the CINECA under the ISCRA IsB28 “HEXTIM”, IsCc2 "SFERA" and IsCa7 "MUSCHI2D" projects.
\medskip

\appendix
\renewcommand{\thesection}{S\arabic{section}}

\renewcommand{\thefigure}{S\arabic{figure}} 
\renewcommand{\thetable}{S\arabic{table}}

\section*{Supporting Information} 
\addcontentsline{toc}{section}{Supplementary Information}

\setcounter{figure}{0} 
\setcounter{table}{0}  
\setcounter{section}{0} 

\section{Symmetry mode analysis for the AFE structure}
\begin{figure}[h]
\centering
\includegraphics[width=10.5 cm,angle=0]{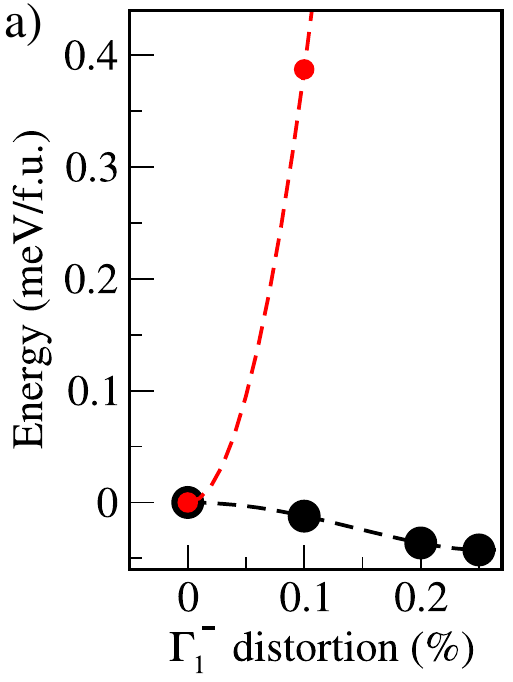}
\caption{
Energy as a function of AFE $\Gamma_1^-$ distortion, calculated by using the GGA (red line) and HSE (black line). 
}\label{AFE_SMA}
\end{figure}

The symmetry mode analysis reveals that the $\Gamma_1^-$ mode is the primary distortion driving the transition from the PE to the AFE phase in SbSI. Starting from the centrosymmetric PE Pnam structure, we introduce various amplitudes of the $\Gamma_1^-$ distortion, keeping the lattice parameters fixed, and compute the total energies of the resulting structures.

Our HSE calculations show a clear energy gain relative to the PE phase as the $\Gamma_1^-$ distortion is condensed, thus providing a compelling energetic justification for the structural transition to the AFE phase. In contrast, analogous calculations performed using the standard GGA functional indicate an energy increase, which fails to stabilize the AFE structure.

\section{Structural properties}

Tab. \ref{Parameters} reports the lattice constants and the atomic Wyckoff positions optimized using the HSE functional and taking into account the inter-chain Van der Waals (vdW) dispersion energy by means of the DFT-D3 correction. These results are also compared with the corresponding experimental data. 

There is an overall good agreement between the calculated and experimental results for the FE phase. In both the FE and AFE phases, the chains elongate along the $c$-axis compared to the PE phase. This displacement of the Sb ions plays a dominant role in driving the transition from the PE to both the FE and AFE phases, as discussed in the main text.

\begin{table}[h]
  \begin{center}
    \caption{Lattice constants and atomic positions of SbSI in the FE (Pna2$_{1}$ space group, 4a Wyckoff positions), AFE (P2$_{1}$2$_{1}$2$_{1}$ space group, 4a Wyckoff positions) and PE (Pnam space group, 4c Wyckoff positions) phases optimized using the HSE functional combined with the DFT-D3 vdW correction and compared with the available experimental data. The atomic Wyckoff positions are indicated in fractional coordinates.}
    \label{tab:table1}
    \begin{tabular}{c|c|c|c|c|}
       & FE HSE & FE Exp. \cite{Oka66} & AFE HSE & PE HSE \\
      \hline
      a(\AA) & 8.56  & 8.49\cite{Donges50} (8.52\cite{Oka66} para) &  8.60 &  8.59\\ 
      b(\AA) & 10.10  & 10.10\cite{Donges50} (10.13\cite{Oka66} para) & 10.07 &  10.13\\
      c(\AA) & 4.16   & 4.16\cite{Donges50} (4.10\cite{Oka66} para)  & 4.17 & 4.09 \\ 
      \hline
     $ x_{Sb}$ & 0.12  & 0.12 & 0.12 & 0.12 \\
      $y_{Sb}$ & 0.12  & 0.12 & 0.12  & 0.12 \\
     $ z_{Sb}$ & 0.31  & 0.30 & 0.19 & 0.25  \\
     $ x_{S}$ & 0.85   & 0.84 & 0.85 & 0.85 \\
      $y_{S}$ & 0.05   & 0.05 & 0.05 & 0.05 \\
     $ z_{S}$ & 0.26  & 0.26 & 0.24 & 0.25  \\
     $ x_{I}$ & 0.50   & 0.51 & 0.50 & 0.50  \\
      $y_{I}$ & 0.83  & 0.83 & 0.83 & 0.82 \\
     $ z_{I}$ & 0.24  & 0.25 & 0.26 & 0.25 \\
    \end{tabular}
    \label{Parameters}
  \end{center}
\end{table}

\clearpage

\section{Single-chain system}
\begin{figure*}
\centering
\includegraphics[width=\textwidth]{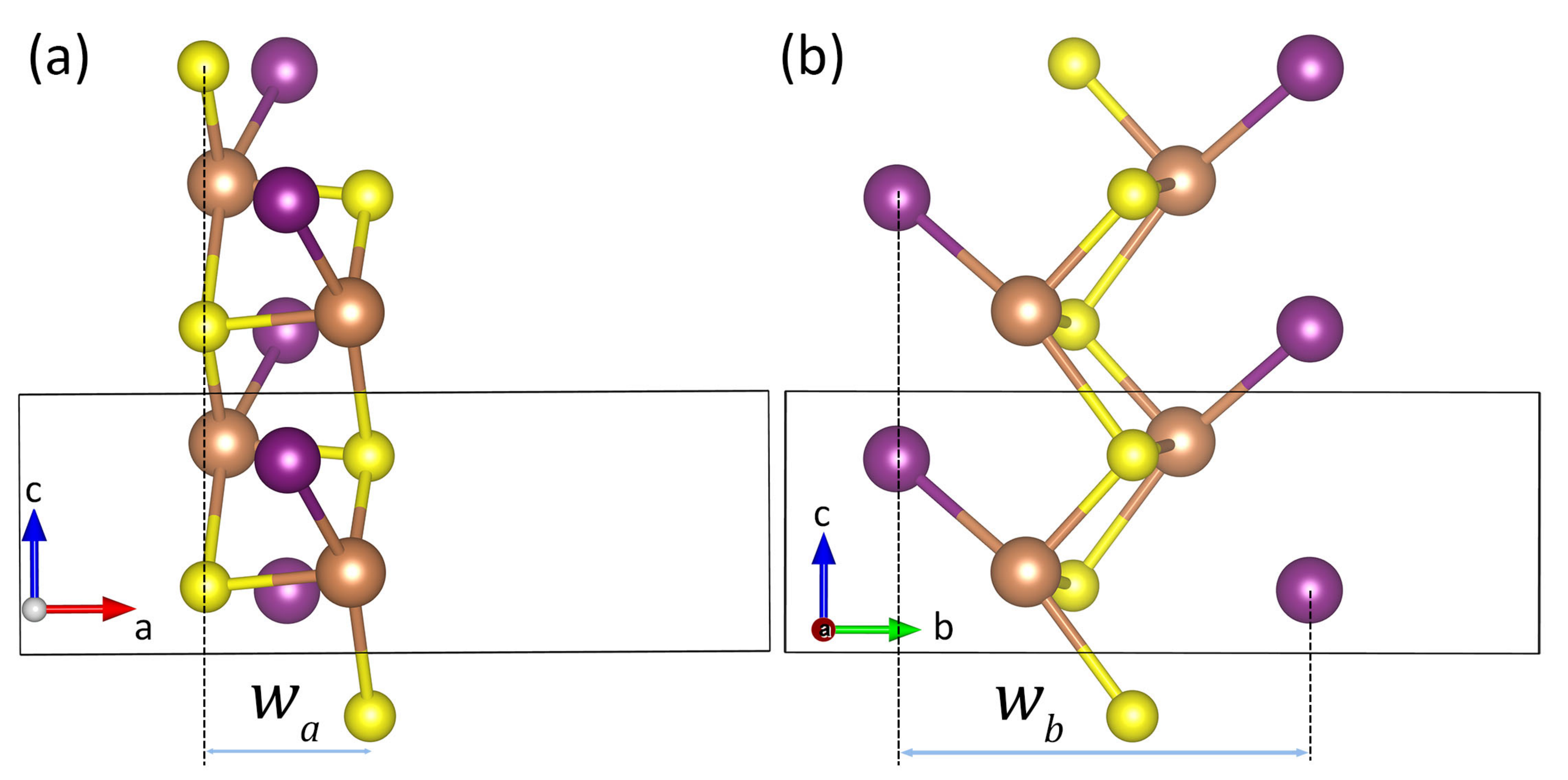}
\caption{Supercell used to model the single-chain system. (a) $a$-$c$ view, (b) $b$-$c$ view. $w_{a}$ and $w_{b}$ are the widths of the chain along the $x$ and $y$ directions, respectively.}\label{structure_SUC}
\end{figure*}

Fig. \ref{structure_SUC} displays the supercell used to model the single-chain system. $w_a$ and $w_b$ are the estimated widths of the chain in the $x$ and $y$ directions, which are used for the renormalization of the photoconductivity in the main text.

\clearpage

\section{Born effective charges (BEC)}
Tables \ref{BEC_SbSI_FE} and \ref{BEC_SbSI_FE_HSE} report the BEC for the FE phase obtained within the GGA and using HSE, respectively. We see that the differences between the HSE and GGA results are overall negligible. 

Table \ref{BEC_SbSI_AFE} presents the GGA BEC for the AFE phase.

\noindent 
\begin{table*}[h]
\begin{centering}
\begin{tabular}{|c|c|c|c|c|c|c|c|c|c|}
\hline 
 & $Z^{*}_{xx}$ & $Z^{*}_{xy}$ & $Z^{*}_{xz}$ & $Z^{*}_{yx}$ & $Z^{*}_{yy}$ & $Z^{*}_{yz}$ & $Z^{*}_{zx}$ & $Z^{*}_{zy}$ & $Z^{*}_{zz}$ \tabularnewline
\hline 
Sb  & 1.96  & 0.07  & 0.03  & 0.87  & 3.51  & 0.38  & 0.13  & -0.12  &   5.97\tabularnewline
\hline
S  & -1.16  &  -0.10  & -0.08  &  -0.85  & -1.72  & -0.09  & -0.27  & -0.40  &  -3.43\tabularnewline
\hline
I & -0.80  &  -0.41  & 0.03  & -0.78  &  -1.79  & -0.17  & -0.06 &  -0.23 & -2.53\tabularnewline
\hline
\end{tabular}
\par\end{centering}
\caption{Born effective charges of the FE phase calculated with the GGA.}
\label{BEC_SbSI_FE} 
\end{table*}

\noindent 
\begin{table*}[h]
\begin{centering}
\begin{tabular}{|c|c|c|c|c|c|c|c|c|c|}
\hline 
 & $Z^{*}_{xx}$ & $Z^{*}_{xy}$ & $Z^{*}_{xz}$ & $Z^{*}_{yx}$ & $Z^{*}_{yy}$ & $Z^{*}_{yz}$ & $Z^{*}_{zx}$ & $Z^{*}_{zy}$ & $Z^{*}_{zz}$ \tabularnewline
\hline 
Sb  & 1.97 & 0.06  & 0.00  & 0.81  & 3.62  & 0.44  & 0.10 & -0.07  &   6.10\tabularnewline
\hline
S  & -1.22  &   -0.12  & -0.08  &   -0.85  & -1.81  & -0.08  &  -0.29  & -0.37  &  -3.56\tabularnewline
\hline
I & -0.75  &   -0.43  & 0.02  &  -0.79  &  -1.81  & -0.20  & -0.08 &  -0.25 & -2.54\tabularnewline
\hline
\end{tabular}
\par\end{centering}
\caption{Born effective charges of the FE phase calculated by using the HSE functional.}
\label{BEC_SbSI_FE_HSE} 
\end{table*}

\noindent 
\begin{table*}[h]
\begin{centering}
\begin{tabular}{|c|c|c|c|c|c|c|c|c|c|}
\hline 
 & $Z^{*}_{xx}$ & $Z^{*}_{xy}$ & $Z^{*}_{xz}$ & $Z^{*}_{yx}$ & $Z^{*}_{yy}$ & $Z^{*}_{yz}$ & $Z^{*}_{zx}$ & $Z^{*}_{zy}$ & $Z^{*}_{zz}$ \tabularnewline
\hline 
Sb  & 1.89  & 0.10  & 0.04  & 0.85  & 3.52  & -0.38  & -0.13 & 0.18  &   6.03\tabularnewline
\hline
S  & -1.14  &   -0.11  & 0.10 &   -0.83  & -1.71  & 0.07  &  0.28  & 0.39 &  -3.45\tabularnewline
\hline
I & -0.75  &   -0.38  & -0.15  &  -0.75  &  -1.81  & -0.22  & -0.24 &  -0.20 & -2.57\tabularnewline
\hline
\end{tabular}
\par\end{centering}
\caption{Born effective charges of AFE phase calculated with the GGA.}
\label{BEC_SbSI_AFE} 
\end{table*}

\clearpage

\section{Band structures}

Fig. \ref{Bands_SbSI} displays the DFT band structure of the FE and AFE phases, alongside the Wannier bands. The DFT bands (red lines) are well reproduced by the Wannier bands (blue lines) in a wide energy range around the Fermi level, namely from -5 to 5 eV. In panel (a), the DFT band structure of the FE phase is obtained within the GGA.
In panel (b), the DFT band structure of the FE phase is obtained using the HSE hybrid functional. 
HSE produces only a rigid shift of the conduction bands compared to GGA without modifying the main electronic features. In panel (c), the DFT band structure of the AFE phase is obtained within the GGA.

\begin{figure}[h]
\centering
\includegraphics[width=5.0cm,angle=270]{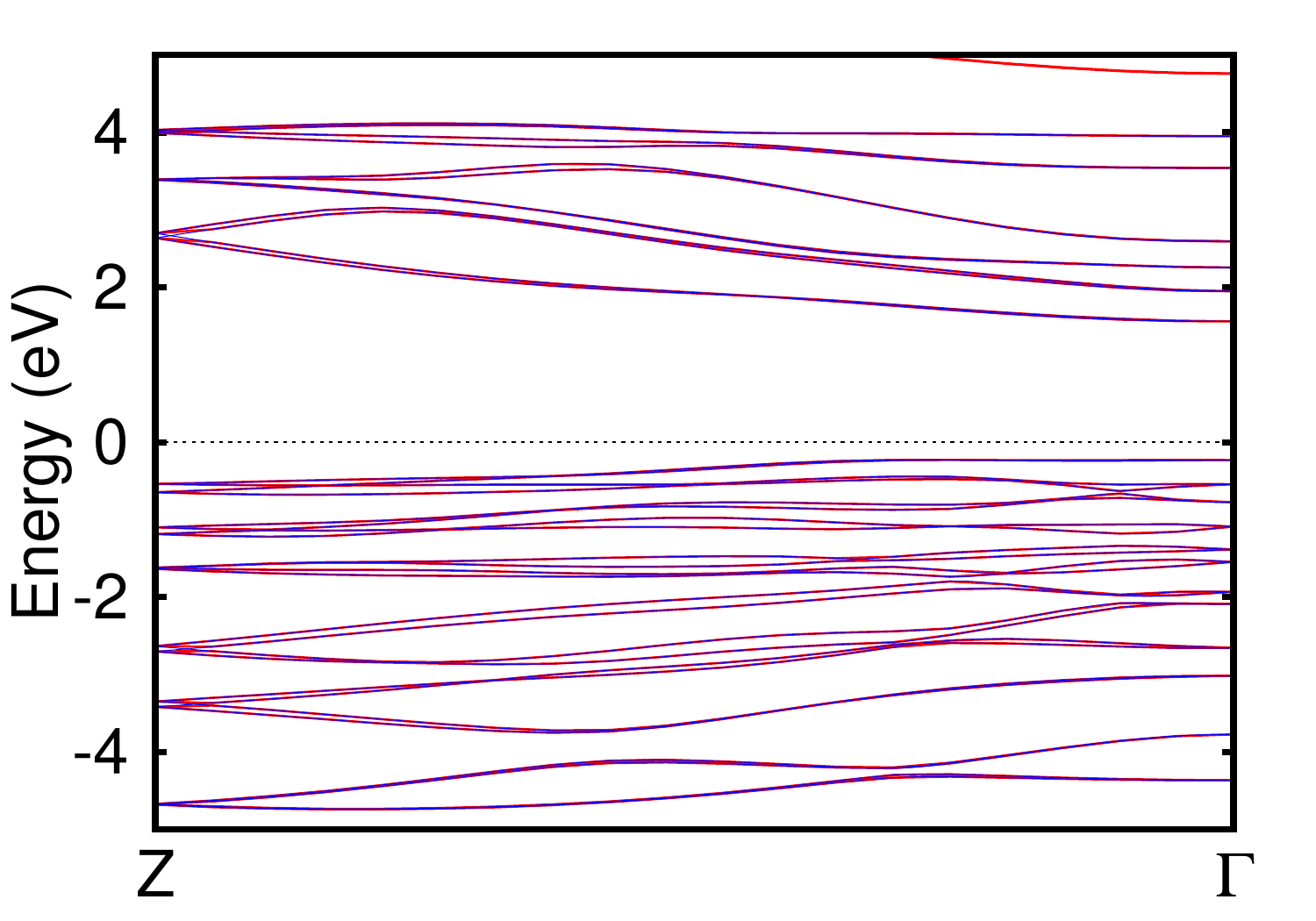}
\caption{DFT band structure (red lines) and  Wannier-interpolated band structure (blue lines) for the single-chain system. The Fermi energy is set at 0 eV. }\label{Bands_SbSI_SUC}
\end{figure}

Fig. \ref{Bands_SbSI_SUC} shows the Wannier band structure for the single-chain system. 

\begin{figure}[h]
\centering
\includegraphics[width=5.0cm,angle=270]{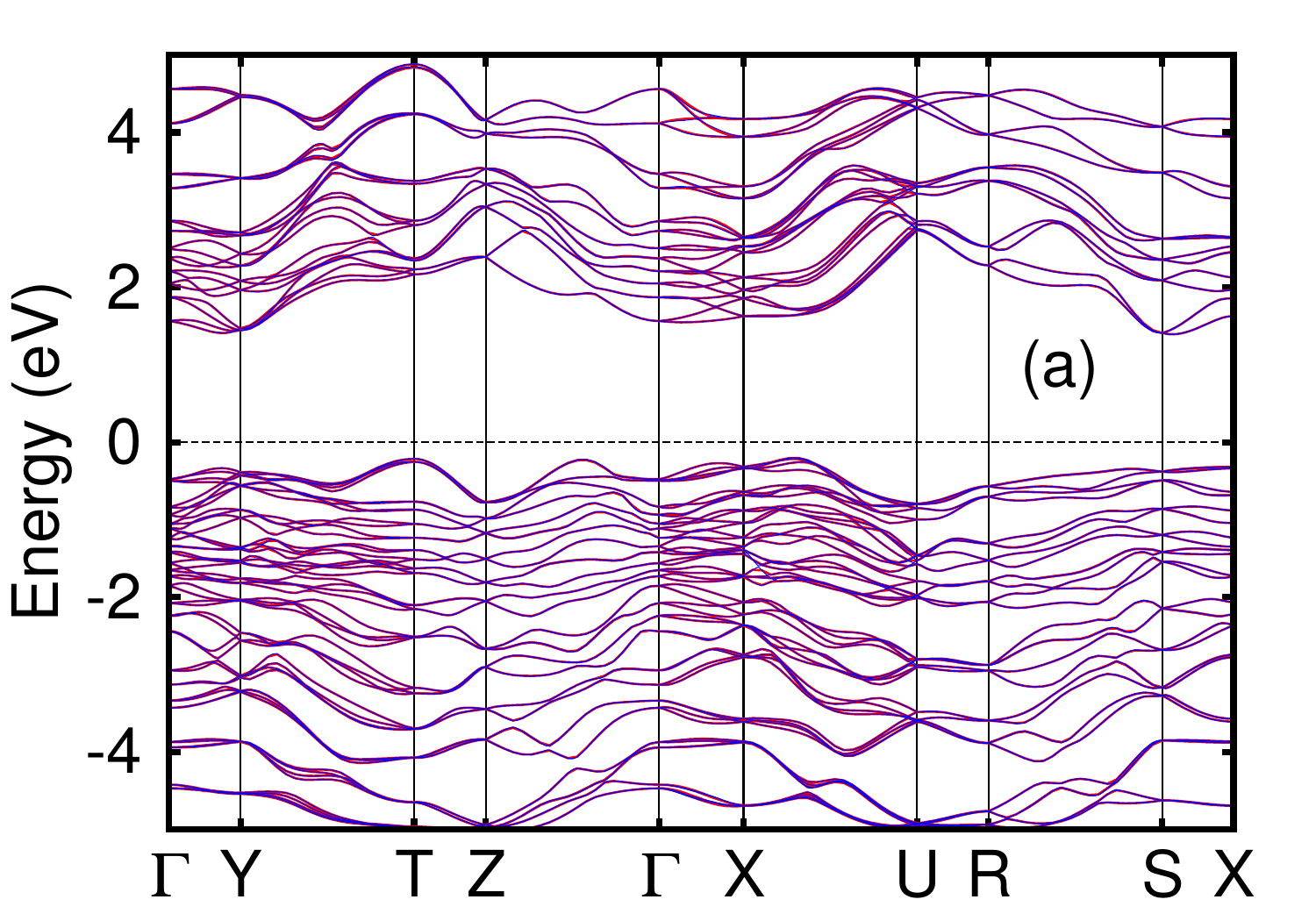}
\includegraphics[width=5.0cm,angle=270]{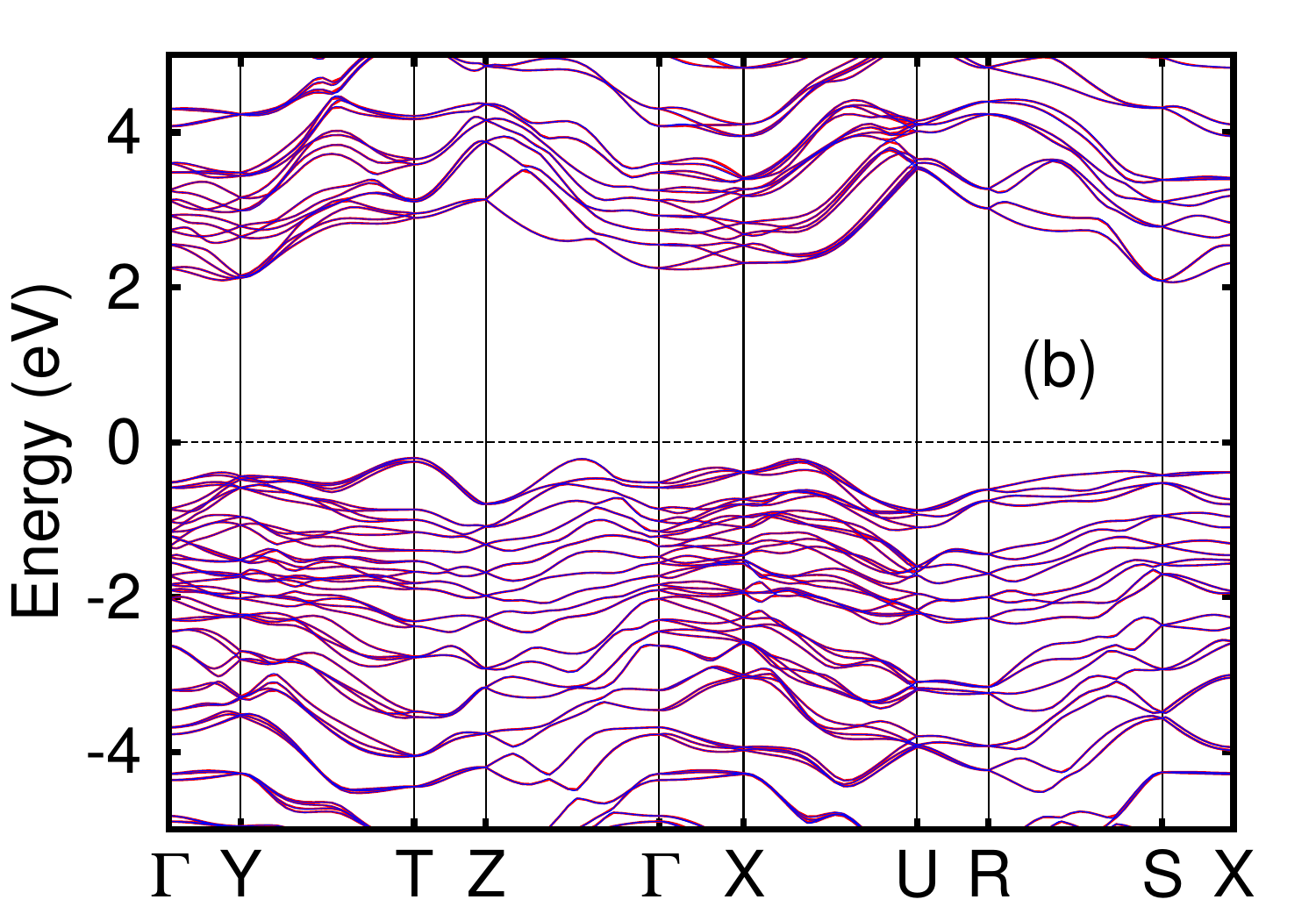}
\includegraphics[width=5.0cm,angle=270]{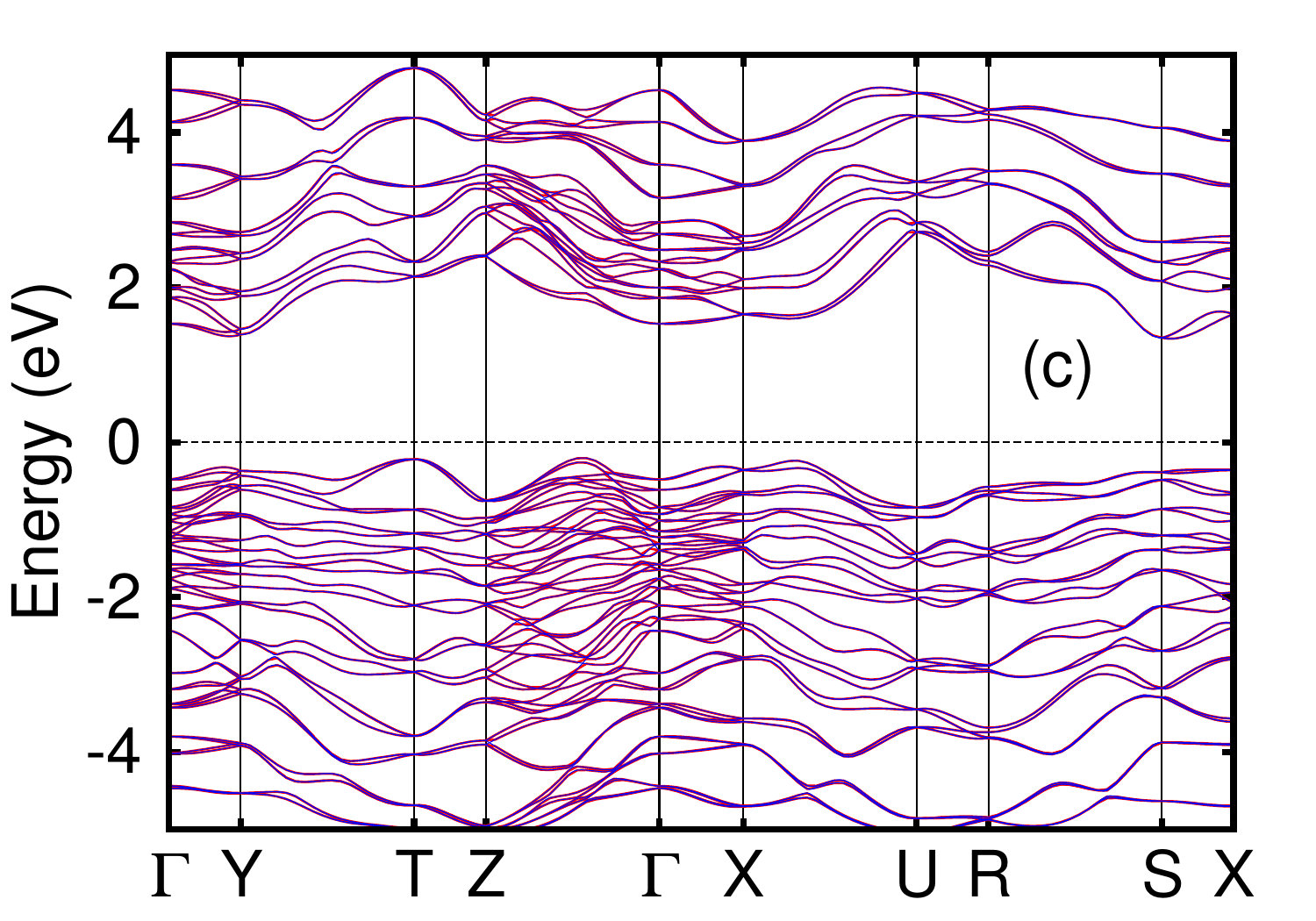}
\caption{DFT band structure (red lines) and Wannier-interpolated band structures (blue lines). (a) DFT Band structure for the FE phase calculated within the GGA. (b) DFT band structure for the FE phase calculated using HSE. (c) DFT Band structure for the AFE phase calculated within the GGA. In all plots, the Fermi energy is set at 0 eV.  }\label{Bands_SbSI}
\end{figure}

\clearpage

\section{Density of states}
\begin{figure}[h]
\centering
\includegraphics[width=5.5cm,angle=270]{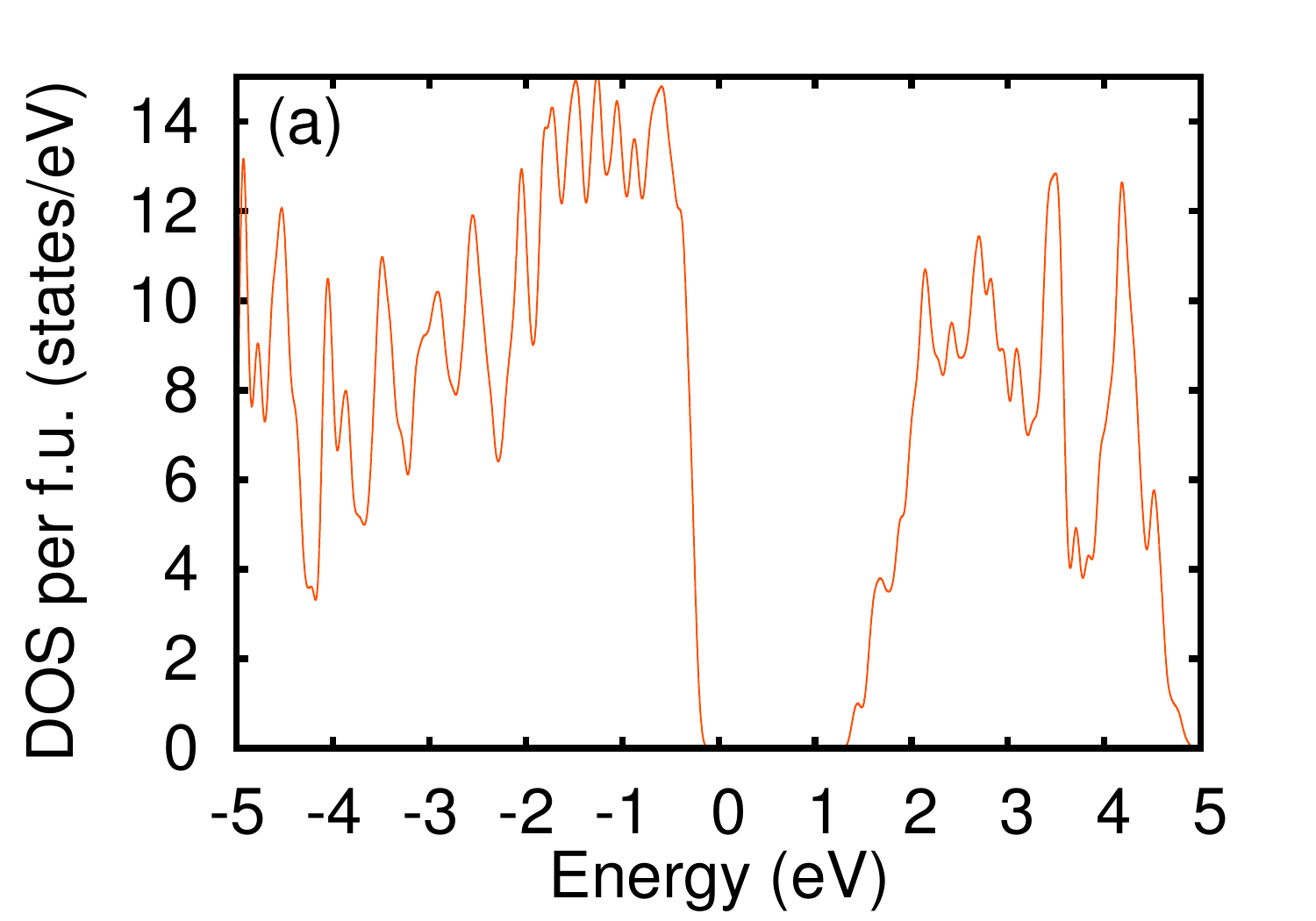}
\includegraphics[width=5.5cm,angle=270]{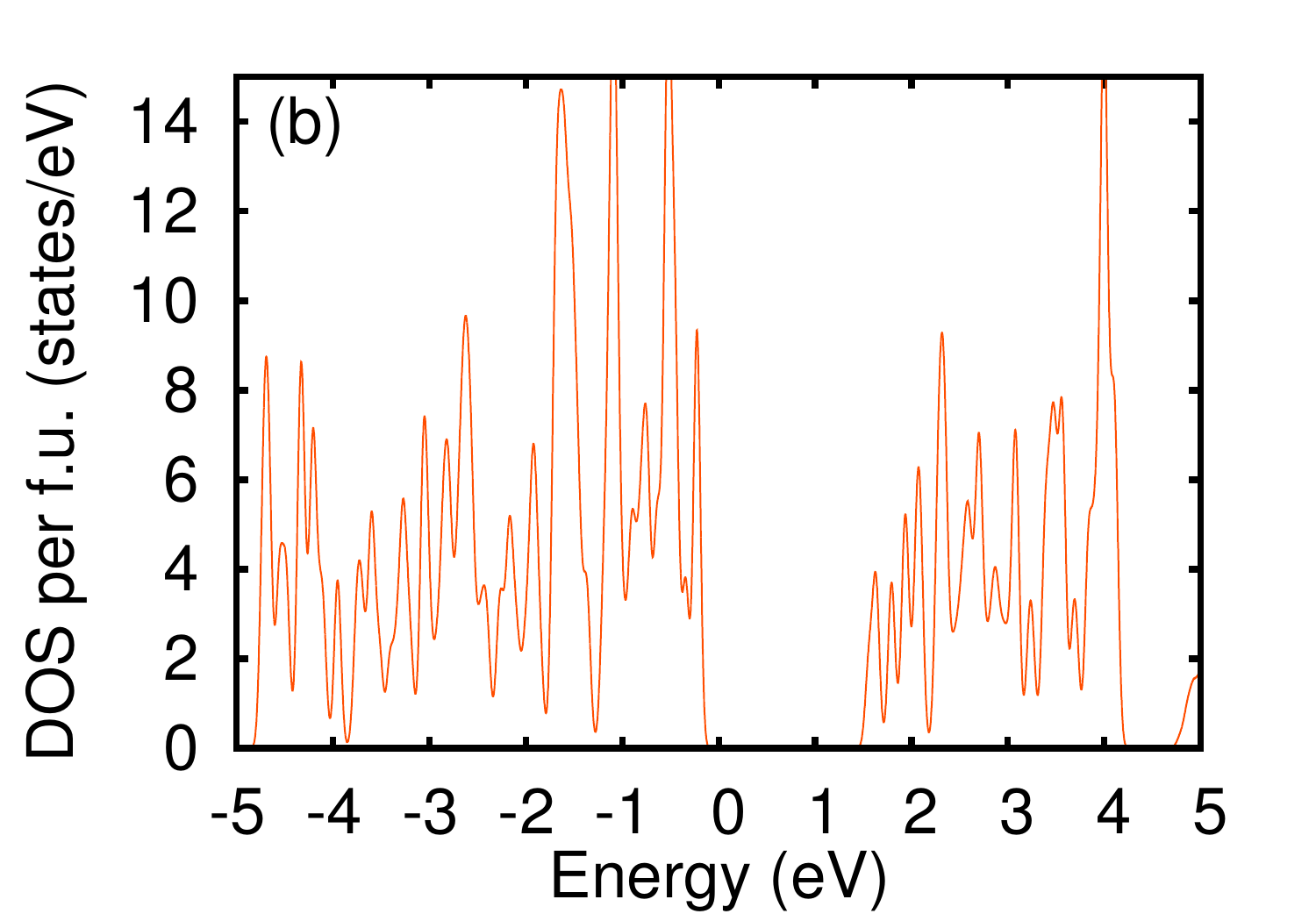}
\includegraphics[width=5.5cm,angle=270]{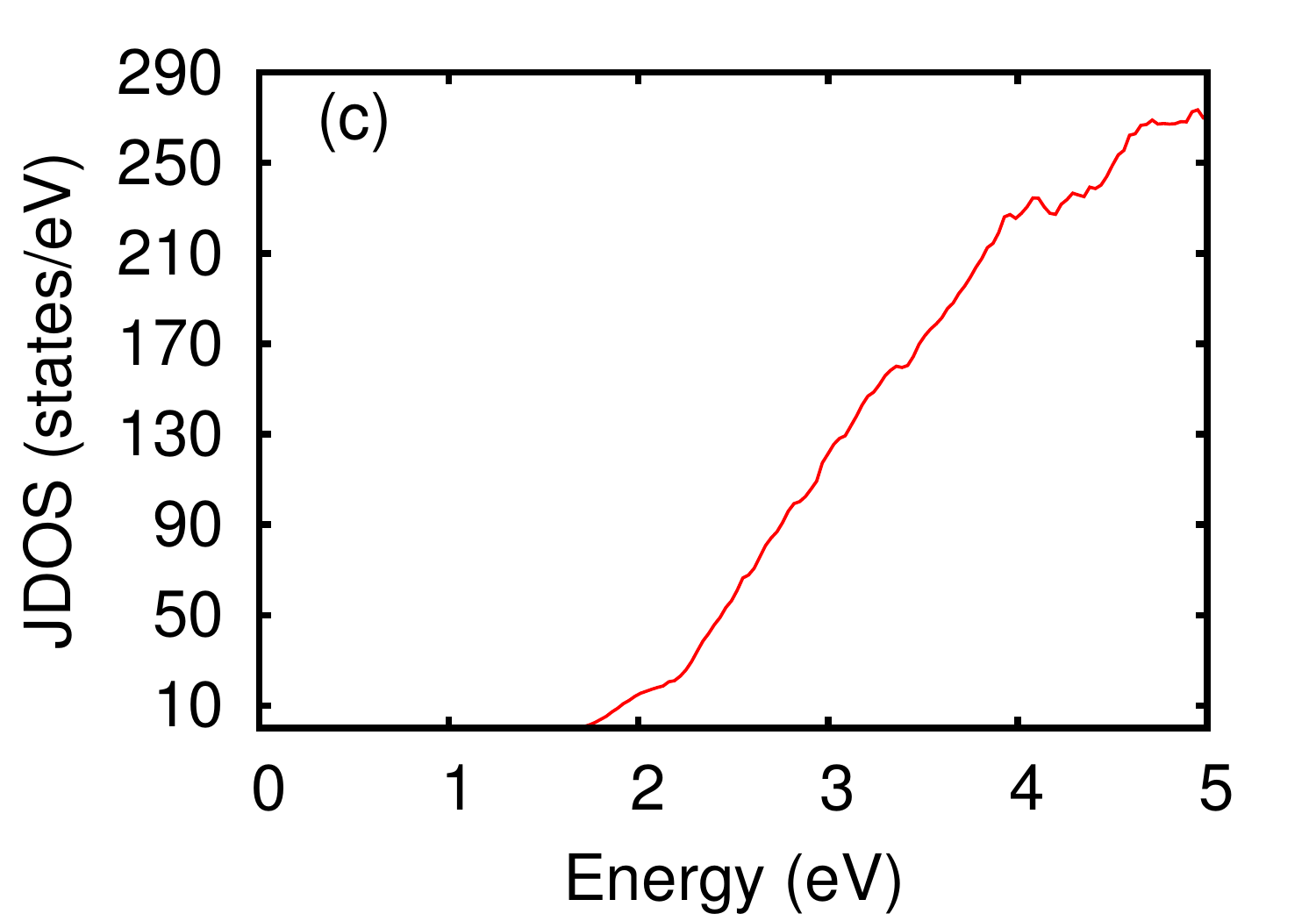}
\includegraphics[width=5.5cm,angle=270]{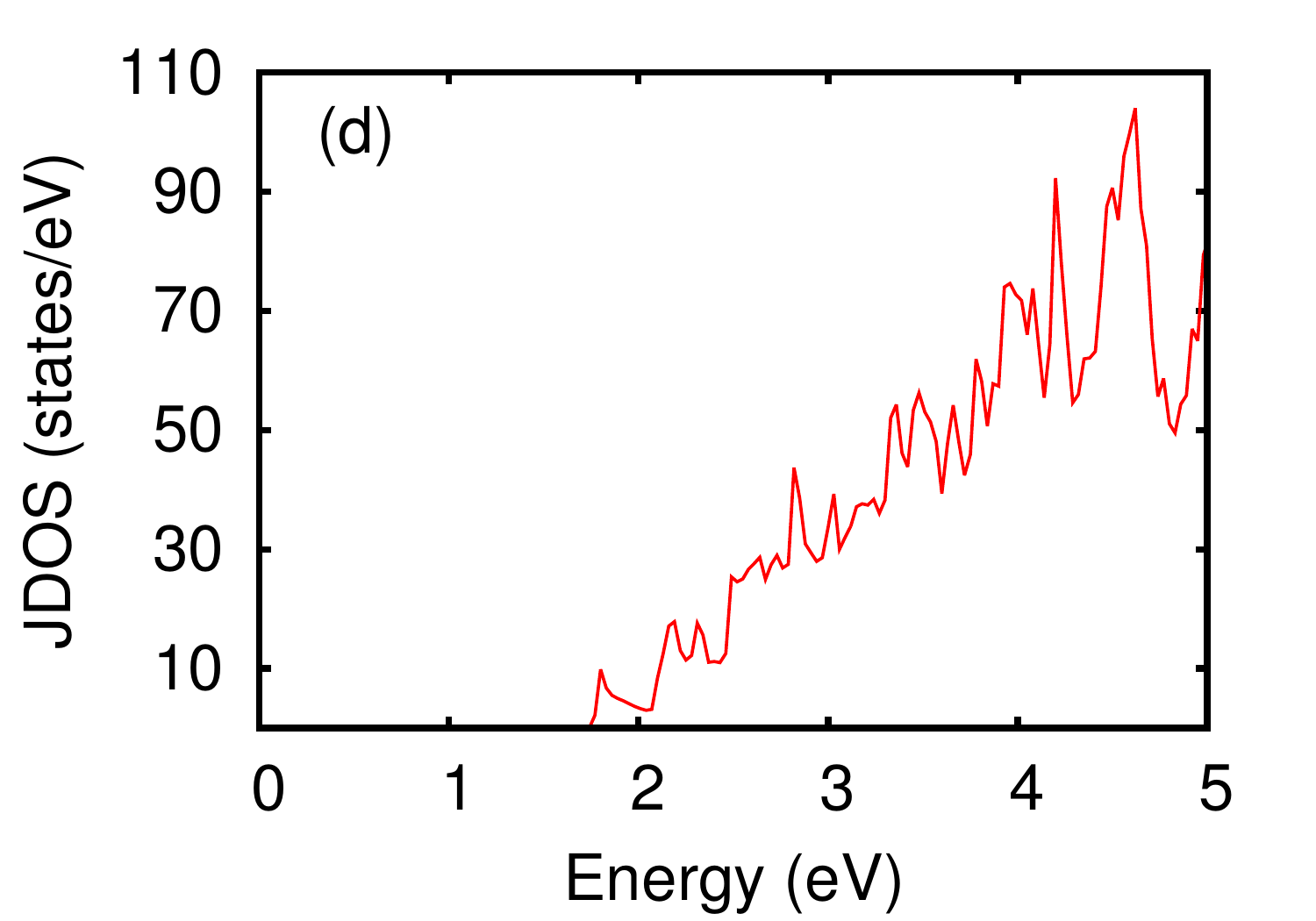}
\caption{DOS of the FE phase for the bulk system (a) and the single-chain system (b). JDOS of the FE phase for the bulk system (c) and the single-chain system (d).}\label{DOS}
\end{figure}

Fig. \ref{DOS}(a) and (b) show the DOS of FE phase for the bulk and the single-chain systems, respectively. The results are obtained with the GGA. In the DOS of the single-chain system, we can clearly observe the appearance of van Hove singularities, which eventually will lead an enhancement of the photoconductivity, as described in the paper.

In addition to the DOS, we can also compute the joint density of states (JDOS). The JDOS quantifies the number of possible optical transitions between the occupied valence bands and the unoccupied conduction bands. It is defined as \cite{Azpiroz18}  

\begin{equation}
D_{joint} = \frac{V_{c}}{\hbar} \int_{BZ} \frac{d^3k}{(2\pi)^3} \sum_{n,m} f_{nm}  \delta(\omega_{mn}-\omega),
\end{equation}

where $V_{c}$ is the cell volume, $n$ and $m$ label two Bloch states $\vert n\rangle$ and $\vert m\rangle$ of energy $\epsilon_n$ and $\epsilon_m$, respectively. $f_{nm} = f_n - f_m$ is the difference between the Fermi occupation factors, and $\omega_{nm} = \left(\epsilon_n - \epsilon_m\right)/\hbar$ is the optical excitation frequency. 

Figs. \ref{DOS}(c) and (d) present the the JDOS of FE phase for the bulk and the single-chain systems, respectively, as calculated within the GGA. 
The onset energy corresponds to the band gap size, which is nearly identical for both the bulk and single-chain systems. Like the DOS, the JDOS of the single chain also displays sharp peaks corresponding to van Hove singularities.

 \clearpage

\section{Bulk photovoltaic effect}




\subsection{Shift photoconductivity}
The shift photoconductivity tensor is defined as \cite{Azpiroz18}

\begin{equation}
\sigma^{abc} = -\frac{i \pi e^3}{4 \hbar^2} \int_{BZ} \frac{d^3k}{(2\pi)^3} \sum_{n,m} f_{nm}[r^b_{mn}r^c_{nm;a} 
+ (b \leftrightarrow c) ] \times [\delta(\omega_{mn}+\omega)+\delta(\omega_{mn}-\omega)].
\end{equation}


In this equation, $e$ is the electron charge and the other quantities are defined as above.
Additionally, we have introduced the interband dipole matrix 
\begin{equation}
r^a_{nm}=(1-\delta_{nm})A^a_{nm},
\end{equation}
and its derivative

\begin{equation}
r^a_{nm;b}=\partial_{b}r^a_{nm} - i (A^b_{nn}-A^b_{mm})r^a_{nm},
\end{equation}
which depend on the Berry connection

\begin{equation}
A^a_{nm}=i \left\langle n | \partial_{a} |m\right\rangle  ,
\end{equation}

with  $\partial_{a}=\frac{\partial}{ \partial k^{a}}$.

\subsection{Injection photoconductivity}
The injection photoconductivity is defined as \cite{Puente23}

\begin{equation}
\eta^{abc} = -\frac{\pi e^3}{\hbar^2} \int_{BZ} \frac{d^3k}{(2\pi)^3} \sum_{n,m} f_{nm}\frac{\partial\omega_{mn}}{\partial k^{a}}r^b_{nm}r^c_{mn}\delta(\omega_{mn}-\omega),
\end{equation}

 which can be separated into a symmetric (S) and an antisymmetric (A) part under the exchange of indices $b \leftrightarrow c$: 

\begin{align*}
\eta^{abc}_\mathrm{S} &= \frac{1}{2}[\eta^{abc}(\omega)+\eta^{acb}(\omega)], \\
\eta^{abc}_\mathrm{A} &= \frac{1}{2}[\eta^{abc}(\omega)-\eta^{acb}(\omega)].
\end{align*}

$\eta^{abc}_\mathrm{S}$ and $\eta^{abc}_\mathrm{A}$ are real and imaginary, respectively. $\eta^{abc}_\mathrm{S}$ describes the response to a linearly polarized light, and vanishes in SbSI because of time-reversal symmetry. $\eta^{abc}_\mathrm{A}$ describes the response to a circularly polarized light. Since, in the paper and in the following, we only consider this antisymmetric part, we drop the subscript ``A'' to keep the notation lighter. 

\subsection{Symmetry analysis for the photoconductivity tensors}

Using the Bilbao Crystallographic Server (BCS) \cite{Bilbao1,Bilbao2}, we can readily identify the non-zero and independent components of the shift and injection photoconductivity tensors based on the space group of our system. 

\subsubsection{Shift photoconductivity}

The shift photoconductivity is a rank-three polar tensor (i.e, even under time-reversal), symmetric with respect to the exchange of its last two indices. By relabeling the Cartesian directions $x$, $y$ and $z$ as $1$, $2$, $3$, respectively, we adopt the BCS abbreviated notation and write its components as $\sigma^{abc}\rightarrow\sigma^{ab}$ where
\[
bc \rightarrow 
\begin{cases} 
b & \text{if } b = c \\
9 - (b + c) & \text{if } b \neq c
\end{cases}.
\]
The independent shift photoconductivity components can then be presented in a table with three rows and six columns, corresponding to the indices $a$ and $b$, respectively.

For the FE phase (space group Pna2$_1$), we obtain
\begin{table}[h]
\centering
\begin{tabular}{|c|c|c|c|c|c|c|}
\hline
$\sigma^{ab}$ & 1 & 2 & 3 & 4 & 5 & 6 \\ 
\hline
1 & 0 & 0 & 0 & 0 & $\sigma^{15}$ & 0 \\ 
\hline
2 & 0 & 0 & 0 & $\sigma^{24}$ & 0 & 0 \\ 
\hline
3 & $\sigma^{31}$ & $\sigma^{32}$ & $\sigma^{33}$ & 0 & 0 & 0 \\ 
\hline
\end{tabular}
\label{symmetries_Pna21}
\end{table}

\noindent where the non-zero components are $\sigma^{15}\rightarrow\sigma^{113}\equiv\sigma^{xxz}$,
$\sigma^{24}\rightarrow\sigma^{223}\equiv\sigma^{yyz}$,
$\sigma^{31}\rightarrow\sigma^{311}\equiv\sigma^{zxx}$,
$\sigma^{32}\rightarrow\sigma^{322}\equiv\sigma^{zyy}$, and 
$\sigma^{33}\rightarrow\sigma^{333}\equiv\sigma^{zzz}$,
as reported in the main text.

In contrast, for the AFE phase (space group P2$_1$2$_1$2$_1$), we obtain
\begin{table}[h]
\centering
\begin{tabular}{|c|c|c|c|c|c|c|}
\hline
 & 1 & 2 & 3 & 4 & 5 & 6 \\ 
\hline
1 & 0 & 0 & 0 & $\sigma^{14}$ & 0 & 0 \\ 
\hline
2 & 0 & 0 & 0 & 0 & $\sigma^{25}$ & 0 \\ 
\hline
3 & 0 & 0 & 0 & 0 & 0 & $\sigma^{36}$ \\ 
\hline
\end{tabular}
\label{symmetries_P212121}
\end{table}

\noindent where the non-zero components are $\sigma^{14}\rightarrow\sigma^{123}\equiv\sigma^{xyz}$,
$\sigma^{25}\rightarrow\sigma^{213}\equiv\sigma^{yxz}$, and
$\sigma^{36}\rightarrow\sigma^{312}\equiv\sigma^{zxy}$.

\subsubsection{Circular injection photoconductivity}

The circular injection photoconductivity is a rank-three polar tensor (i.e, even under time-reversal), antisymmetric with respect to the exchange of its last two indices. Following the BCS, its components $\eta^{abc}$ can be represented in a table with three rows, corresponding to the index $a$, and nine columns corresponding to the indices $bc$. 

For the FE phase (space group Pna2$_1$), we obtain


\begin{table}[h]
\centering
\begin{tabular}{|c|c|c|c|c|c|c|c|c|c|c|}
\hline
 $\eta^{abc}$& 11 & 21 & 31 & 12 & 22 & 32 & 13 & 23 & 33 \\ \hline
 1 & 0 & 0 & $-\eta^{113}$ & 0 & 0 & 0 & $\eta^{113}$ & 0 & 0 \\ \hline
 2 & 0 & 0 & 0 & 0 & 0 & $-\eta^{223}$ & 0 & $\eta^{223}$ & 0 \\ \hline
 3 & 0 & 0 & 0 & 0 & 0 & 0 & 0 & 0 & 0 \\ 
\hline
\end{tabular}
\label{IC_symmetries_Pna21}
\end{table}
\noindent Thus, there are four non-zero components of the circular injection photoconductivity, but only two of them are independent, namely $\eta^{113}\equiv\eta^{xxz}$ and
$\eta^{223}\equiv\eta^{yyz}$, as reported in the main text.

In the case of the AFE phase (space group P2$_1$2$_1$2$_1$), we obtain

\begin{table}[H]
\centering
\begin{tabular}{|c|c|c|c|c|c|c|c|c|c|c|}
\hline
$\eta^{abc}$ & 11 & 21 & 31 & 12 & 22 & 32 & 13 & 23 & 33 \\ 
\hline
1 & 0 & 0 & 0 & 0 & 0 & $-\eta^{123}$ & 0 & $\eta^{123}$ & 0 \\ 
\hline
2 & 0 & 0 & $-\eta^{213}$ & 0 & 0 & 0 & $\eta^{213}$ & 0 & 0 \\ 
\hline
3 & 0 & $-\eta^{312}$ & 0 & $\eta^{312}$ & 0 & 0 & 0 & 0 & 0 \\ 
\hline
\end{tabular}
\label{IC_symmetries_P212121}
\end{table}

\noindent We therefore find three independent components of the circular injection photoconductivity, that are $\eta^{123}\equiv\sigma^{xyz}$, 
$\eta^{213}\equiv\sigma^{yxz}$, and $\eta^{321}\equiv\sigma^{zxy}$.


\clearpage

\section{Photoconductivity: GGA versus HSE}

\begin{figure}[h]
\centering
\includegraphics[width=6.5cm,angle=270]{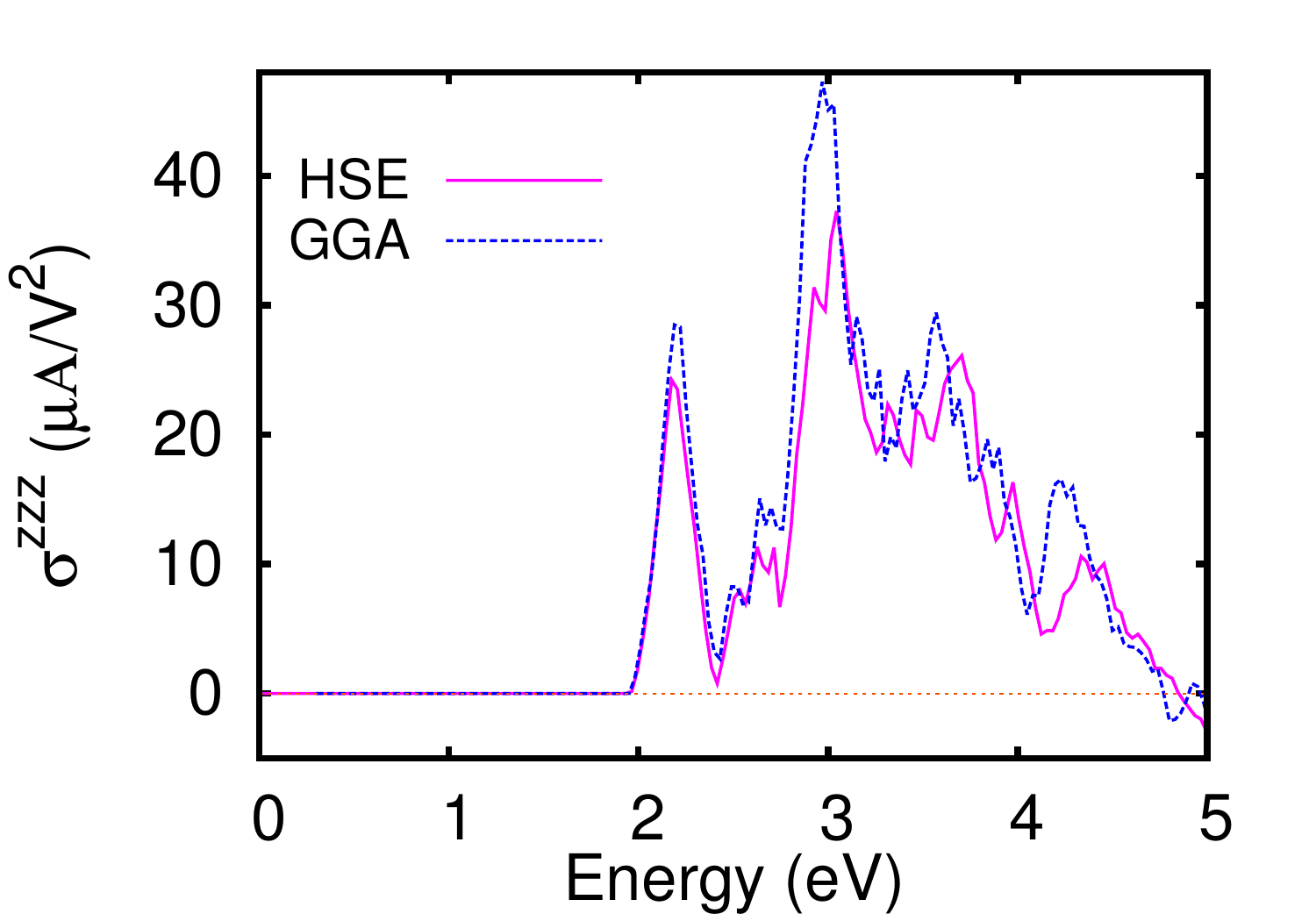}
\caption{$\sigma^{zzz}$ component of the shift photoconductivity tensor calculated within the GGA and using HSE.}\label{SHC_HSE_GGA}
\end{figure}

Fig. \ref{SHC_HSE_GGA} compares the $\sigma^{zzz}$ component of the shift photoconductivity tensor for the FE phase calculated within the GGA and using HSE. Both plots are shifted to align the onset energies with the experimental band gap. We observe that the peak at about 3 eV is slightly reduced in the HSE result. Nonetheless, the differences between the two spectra are overall minor, reflecting the negligible differences in the band structure obtained with the two methods. Thus, we present only the GGA results in the main text.

\clearpage



\bibliographystyle{unsrt}
\bibliography{biblio}

\end{document}